\def\Re{{\rm Re}}
\def\Im{{\rm Im}}
\def\PiL{\Pi_{\rm L}}
\def\PiT{\Pi_{\rm T}}

\def\x{{\bf x}}
\def\p{{\bf p}}
\def\q{{\bf q}}
\def\k{{\bf k}}
\def\v{{\bf v}}
\def\u{{\bf u}}
\def\E{{\bf E}}
\def\B{{\bf B}}

\def\C{{\cal C}}

\def\B{{\rm B}}

\def\naBla{{\bf \nabla}}
\def\ij{{i \cdots j}}

\def\ca{C_{\rm A}}
\def\cf{C_{\rm F}}
\def\da{d_{\rm A}}
\def\ta{T_{\rm A}}
\def\tf{T_{\rm F}}
\def\df{d_{\rm F}}
\def\nf{N_{\rm f}}
\def\nc{N_{\rm c}}
\def\mD{m_{\rm D}}

\def\trans{\top}

\newcommand\ansatz{{\it Ansatz}}

\newcommand\Eq[1]{Eq.~(\ref{#1})}
\newcommand{\nott}[1]{\not{\! #1 \,}}

\def\half{{\textstyle{1\over2}}}

\def\MSbar{$\overline{\hbox{MS}}$}

\def\alphas{\alpha_{\rm s}}

\def\ttuuss{\;\raisebox{0.8em}{\mbox{``}} \! \! \left( \frac{t^2+u^2}{s^2}\right) \! \!\raisebox{0.8em}{\mbox{''}}\:}
\def\ssuutt{\;\raisebox{0.8em}{\mbox{``}} \! \! \left( \frac{s^2+u^2}{t^2}\right) \! \!\raisebox{0.8em}{\mbox{''}}\:}

\def\gsim{\mbox{~{\raisebox{0.4ex}{$>$}}\hspace{-1.1em}
	{\raisebox{-0.6ex}{$\sim$}}~}}
\def\lsim{\mbox{~{\raisebox{0.4ex}{$<$}}\hspace{-1.1em}
	{\raisebox{-0.6ex}{$\sim$}}~}}

\newfam\scrfam                                          
\global\font\twelvescr=rsfs10 scaled\magstep1%
\global\font\eightscr=rsfs7 scaled\magstep1%
\global\font\sixscr=rsfs5 scaled\magstep1%
\skewchar\twelvescr='177\skewchar\eightscr='177\skewchar\sixscr='177%
\textfont\scrfam=\twelvescr\scriptfont\scrfam=\eightscr
\scriptscriptfont\scrfam=\sixscr

\documentstyle[preprint,aps,fixes,epsf,eqsecnum]{revtex}

\makeatletter           
\@floatstrue
\def\figure{\let\@capwidth\columnwidth\@float{figure}}
\let\endfigure\end@float
\@namedef{figure*}{\let\@capwidth\textwidth\@dblfloat{figure}}
\@namedef{endfigure*}{\end@dblfloat}
\def\table{\let\@capwidth\columnwidth\@float{table}}
\let\endtable\end@float
\@namedef{table*}{\let\@capwidth\textwidth\@dblfloat{table}}
\@namedef{endtable*}{\end@dblfloat}
\def\la{\label}
\makeatother

\tightenlines
\newcommand\pcite[1]{\protect{\cite{#1}}}

\begin {document}


\preprint {UW/PT 01--11}

\title
    {
	Transport Coefficients in Large $N_f$ Gauge Theory:  \\
	Testing Hard Thermal Loops
    }

\author{Guy D. Moore 
	}
\address
    {%
    Department of Physics,
    University of Washington,
    Seattle, Washington 98195
    }%

\date {April 2001}

\maketitle
\vskip -20pt

\begin {abstract}%
    {%
	We compute shear viscosity and flavor diffusion coefficients for 
	ultra-rel\-a\-tiv\-ist\-ic gauge 
	theory with many fermionic species, $\nf {\gg} 1$, to leading 
	order in $1/\nf$.  The
	calculation is performed both at leading order in the
	effective coupling strength $g^2 \nf$, 
	using the Hard Thermal Loop (HTL) approximation, and completely to 
	all orders in $g^2 \nf$.  This constitutes a nontrivial test
	of how well the HTL approximation works.  We find that
	in this context, the HTL
	approximation works well wherever the renormalization point
	sensitivity of the leading order HTL result is small.
    }%
\end {abstract}

\thispagestyle{empty}

\section {Introduction}
\la{sec:intro}

Many problems in early universe cosmology, and in heavy ion physics, require
thermal field theory.  Perturbation theory was developed for thermal
field theory decades ago \cite{Keldysh}.  However, the application of
perturbative techniques to problems
sensitive to long time scales, or to low frequency or wave number plasma
excitations, runs into subtleties which have made it very difficult to
compute interesting time dependent phenomena even at leading order in
the gauge coupling.

The physics of Fourier modes with frequency and
wave number $(\omega , k) \sim gT$ (with $g$ the gauge coupling and $T$
the temperature), the so called ``soft'' degrees of freedom, cannot be
treated even at leading order in $g$ without the resummation of a class
of diagrams, called the Hard Thermal Loops (HTL's) \cite{HTL}.  
This problem was elucidated by Braaten, Pisarski, Frenkel, Taylor, and
Wong, who showed that, once the HTL diagrams have been re-summed into
the propagators and vertices of soft excitations, 
one may make a loopwise expansion which is now an expansion in $g$.
This breakthrough has made possible the calculation at leading order 
of a number of time dependent plasma properties, such as particle
damping rates \cite{damprate} and hard particle energy loss
\cite{energyloss}.
It has also allowed the development of an effective
theory for even more infrared fields, relevant to baryon number
violation in the standard model \cite{bodeker}.

Another problem of interest in thermal field theory is to compute very
long time scale properties of the plasma.  Since the most infrared
properties of the plasma admit a hydrodynamic description, this requires,
besides understanding of the thermodynamics, the calculation of 
transport coefficients (shear and bulk
viscosity, electric conductivity, fermionic number diffusion).
The transport coefficients
can be formally
related to {\em zero} frequency and momentum limits of correlation
functions of physical observables, and require for their computation,
at leading
order in the coupling, the resummation of an infinite class of diagrams,
which for scalar field theory has been shown to be ladder graphs
\cite{Jeon}.

To date, the only leading order calculation of a
transport coefficient in relativistic field theory is for one component
scalar field theory \cite{Jeon}.  Despite a substantial literature
\cite{HosoyaKajantie,Hosoya_and_co,relax1,relax2,relax3,relax4,%
BMPRa,Heiselberg,Heiselberg_diff,BaymHeiselberg,JPT1,MooreProkopec,JPT2},
it is only recently that transport coefficients  have been computed
correctly in a gauge theory to leading order in the {\em logarithm} of
the coupling \cite{paper1}.  This is partly because there are new
complications in hot gauge theories; unlike scalar field theory,
hydrodynamic transport coefficients are sensitive at leading order to
soft physics\footnote{%
    For the case of bulk viscosity, even in the scalar theory there is
    sensitivity to soft physics \pcite{Jeon}.  Bulk viscosity is 
    more complicated for other reasons, and will not be considered
    further here.
    }.
Hence their calculation requires control of both long time scales and
soft momenta.  

Clearly it would be valuable to compute transport coefficients to full
leading order in the coupling within a gauge theory, even if only in
some simplifying regime which rendered the calculation easier.  Such a
calculation will require resummation of hard thermal loops.  It
would be even more valuable to compute transport coefficients {\em
beyond} leading order in the coupling.  Aside from the intrinsic
interest of the calculation, such a calculation would also
permit a test of the quality of the HTL approximation.  The best measure
of the usefulness of a perturbative expansion is to see how fast the
series converges as one goes to higher and higher order.  So far, in
applications where HTL resummation is necessary, even the best results are
at leading order in the coupling\footnote{Thermodynamic quantities are a
notable exception, where for instance the pressure is known to fifth
order in the coupling \pcite{BraatenNieto}.}.  Therefore it is hard to
know what confidence to assign to a result computed at leading order
using hard thermal loops.

The purpose of this paper is to compute transport coefficients, at
leading order and to all orders in the coupling,
in a toy theory where it is possible to perform the calculation.
The toy theory is QED or SU($\nc$) QCD, with a
large number $\nf$ of massless fermions, $\nf \gg \nc$, $\nf \gg 1$.  We treat
$g^2$ as very small, and will work only to leading nontrivial order in
this quantity; but the t'Hooft-like coupling $g^2 \nf$ can either be
expanded in perturbatively, or treated as $O(1)$.  As we will see, at
leading order in $1/\nf$, nonabelian effects are unimportant, so we will
generally use language and normalizations of QED.

QED with $\nf \gg 1$ is still a very nontrivial
theory, with substantial similarities to realistic QED or QCD.  Unlike a
scalar theory, the theory is derivative coupled.  HTL
corrections are needed for soft gauge boson lines, and represent a rich
set of plasma physics phenomena, such as Debye screening and Landau
damping.  At weak coupling, scattering cross sections show the Coulombic
divergence, cut off by screening effects; there is also interesting
collinear physics with an analog in full QED or QCD (see
Sec.~\ref{sec:schann}).  Yet the theory is enough simpler that we can
treat it to all orders in $g^2 \nf$.  Naturally, this is because it is
missing some of the physics of QED or QCD.  For instance, bremsstrahlung is
absent at leading order in $1/\nf$, even at large $g^2 \nf$.  The added
complications in QED and QCD may mean that the HTL expansion is worse
behaved in those theories, so our results on the convergence should be
considered as optimistic estimates for the series convergence in
realistic theories.

The large $\nf$ theory has a peculiar structure; while there are an
enormous number, $O(\nf)$, of fermionic degrees of freedom, there are
only $O(1)$ gauge field degrees of freedom (for QCD we treat $\nc \sim
1$).  Since the vertices of the theory always involve the gauge boson,
the rate at which a fermion interacts is actually very small; there are
only $O(1)$ degrees of freedom for a fermion to couple to, and $g^2 \sim
1/\nf$ is very small.  The gauge bosons, on the other hand, couple to
all of the different fermions.  Though the interactions are individually
weak, $g^2 \nf$ is not small, so for the gauge bosons the interactions
are significant.  

\begin{figure}[t]
\centerline{
\epsfxsize=2in\epsfbox{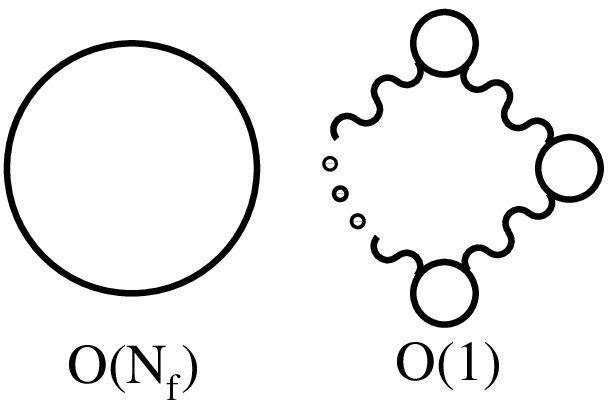} \hspace{0.5in}
\epsfxsize=3in\epsfbox{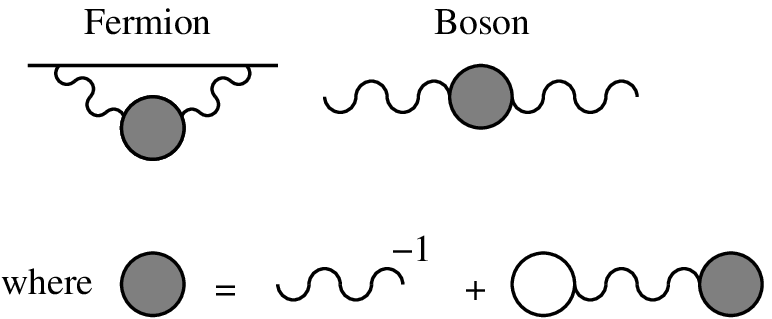}
}
\caption{All bubble graphs, left, and the resulting self-energies
when they are cut, right. \label{fig:bubbles}}
\end{figure}

This structure leads to key simplifications which make the calculation
of transport coefficients tractable without an expansion in $g^2 \nf$ small.  
Since the fermions only couple to $O(1)$ gauge boson degrees of freedom,
the fermionic self-energy is $O(g^2) \sim
O(1/\nf) \ll 1$.  This behavior is clear at one loop, but it is also
true to all orders in loops.  
The easiest way to see this is to draw all bubble diagrams which appear
in the theory at $O(\nf^0)$, which are illustrated in
Fig.~\ref{fig:bubbles}.  Since any self-energy can be generated by
cutting one line in a bubble diagram, the fermionic self-energy to the
order of interest can be obtained from this set of bubble diagrams.

The smallness of the fermionic self-energy means that fermions are
almost free particles.  In particular, the imaginary part of the
fermionic self-energy is $\sim 1/\nf$ for all momenta. Therefore, there
is a sharp quasi-particle mass shell for fermions, even if $g^2
\nf$ is parametrically $O(1)$.  Fermions propagate freely over very long
distances, undergoing occasional collisions with mean separation $\sim
\nf/T$.  For this reason a kinetic theory description
of the fermions is reliable, with $O(g^2 \sim 1/\nf)$ (and therefore
negligible) corrections.  

A kinetic description of fermions only,
without including gauge bosons as kinetic degrees of freedom, is also
sufficient.  This is so because fermions dominate the observables of
interest in this paper, the stress-energy $T_{\mu \nu}$ and all
fermionic number currents, simply because there are $O(\nf)$ times as
many fermionic as gauge degrees of freedom.

The other key simplification is that, at leading order, the fermion
and gauge boson self-energies each have a very simple structure.  The
gauge boson propagator is, at $O(\nf^0)$ but to all orders in $g^2 \nf$,
given by re-summing a one loop fermionic self-energy,
\begin{equation}
D_{\mu \nu}^{-1} = D_{\mu \nu,{\rm free}}^{-1} 
	- \Pi_{\mu \nu}({\rm 1 \; loop}) \, .
\label{eq:D_inv}
\end{equation}
This should be clear from Fig.~\ref{fig:bubbles}.
The fermionic self-energy is gotten by putting the resummed gauge boson
propagator into the self-energy diagram shown in Fig.~\ref{fig:bubbles}.  At
leading order in a $1/\nf$ expansion the fermionic lines needed to
compute $\Pi_{\mu \nu}$ in Eq.~(\ref{eq:D_inv}), and the fermionic line
appearing in the fermionic self-energy, are free theory lines.

These simplifications will be sufficient so that we will be able to solve
for the transport coefficients within kinetic theory, without
uncontrolled approximations.  Our method for solving for the transport
coefficients follows very closely our previous work
\cite{paper1}, except that we will not
need to make a leading log approximation of scattering processes, but
will be able to treat them in a complete way (at leading order in
$1/\nf$).  Some integrals need to be done numerically
and the momentum dependence of the departure from equilibrium has to be
modeled with a several parameter \ansatz, but for both approximations
the error in the treatment can be made arbitrarily small with sufficient
numerical effort; modest effort drives relative errors to $0.1\%$.  

The large $\nf$ theory has a technical problem, which is that it does
not exist, because the gauge boson propagator exhibits a Landau pole at finite
$Q^2$.  However, provided $Q_{\rm Landau} \gg T$ (say, at least $40
T$), this problem is only manifested at energy scales which the thermal
bath probes exponentially rarely.  It is therefore
irrelevant to the thermal physics we study.  Avoiding Landau pole
physics restricts somewhat the range of $g^2 \nf$ we can consider, but it
is not a severe restriction; the perturbative expansion is still not
guaranteed to be well behaved at the largest values of $g^2 \nf$ which
satisfy the above criterion.

An outline of the paper is as follows.  We discuss how the transport
coefficients are to be computed via kinetic theory in
Sec.~\ref{sec:kinetic}.  The most complicated feature of the kinetic
theory is the collision integral, which is discussed in
Sec.~\ref{sec:collision}, with some details relegated to two appendices.
Our exact (at leading order in $1/\nf$) 
results for transport coefficients appear in
Sec.~\ref{sec:results}.  Then, Sec.~\ref{sec:leading} discusses how the
HTL approximation can be used in the calculation to obtain the result to
leading order in $g^2 \nf$, using a simpler scattering 
matrix element.  It presents a comparison
of the leading order result with the exact result.  Finally, there is a
conclusion, Sec.~\ref{sec:conclusion}.  A very brief summary of the
conclusion is that, at relatively weak coupling, the leading order
calculation works extremely well; but at couplings relevant for even the
highest energy heavy ion collisions currently being considered, a full
leading order treatment is at best a factor of 2 estimate, because of
severe renormalization scale uncertainty.  The renormalization point
suggested by dimensional reduction turns out to do surprisingly well at
large $\nf$; but it is difficult to extrapolate, whether this will occur
more generally.

\section{Transport Coefficients and Kinetic Theory}
\label{sec:kinetic}

\subsection{Transport coefficients}

The equilibrium state of a plasma, or any system, is determined by the
densities of all conserved quantities.  For the theory we consider,
massless QED or QCD with many Dirac fermion species, the conserved
quantities are the energy-momentum tensor, fermionic particle number,
and the charges under 
SU($\nc$)$\times$SU($\nc$) flavor rotations.  In the grand canonical
ensemble these are determined by the timelike inverse temperature
4-vector $\beta_\mu$ and various chemical
potentials $\mu^s$ (with $s$ a combined species and spin label).  Here
$\beta_\mu$ determines both the temperature $1/T = \beta =
\sqrt{-\beta_\mu \beta^\mu}$ and the velocity of flow $u_i =
\beta_i / \beta_0$ of the fluid, and the $\mu^s$ determine the values of
the conserved particle numbers.  [Note that we use a $({-}{+}{+}{+})$ index
metric throughout, so timelike 4-vectors have negative squares.]

The hydrodynamic regime is the regime where the densities of conserved
quantities vary slowly enough in space that the local state at each
point can be described up to small corrections by the equilibrium
configuration, but with spacetime dependent $\beta_\mu$ and $\mu^s$.
Transport coefficients are defined in terms of the departure from
equilibrium, due to the gradients of conserved quantities, and are
related to entropy production and the damping away of the spatial
inhomogeneity.  In particular, viscosities are defined by the difference
between the equilibrium and actual stress energy tensor.  If in
equilibrium and in the (local) rest frame $\u(x)=0$ the stress tensor
is given by $T_{ij} = \delta_{ij} {\cal P}$ 
(${\cal P}$ the pressure, a function of $\beta_\mu$ and the $\mu^s$),
then when $\u$ has gradients, $T_{ij}$ is given to lowest order in the
gradients of $\u$ by
\begin{equation}
T_{ij} = \delta_{ij} {\cal P} - \eta \Big[ \nabla_i u_j + 
	\nabla_j u_i - \frac{2}{3} \delta_{ij} \nabla_l u_l
	\Big] - \zeta \delta_{ij} \nabla_l u_l \, ,
\end{equation}
with $\eta$ the shear viscosity and $\zeta$ the bulk viscosity.  (Both
are functions of $\beta$ and the $\mu^s$.)  Evaluating bulk
viscosity is quite complicated because it involves the physics of
particle number changing processes
\cite{Jeon}; we will not treat it here, but estimate
parametrically that $ \zeta \sim \nf^2 T^3$, times a dimensionless
function of $g^2 \nf$, in this theory.

The other transport coefficient we consider is number diffusion.
The diffusion constant for a conserved current $j_\mu$ is defined by the
constitutive relation (in the frame $\u=0$)
\begin{equation}
j_i = - D \nabla_i j_0 \, .
\end{equation}
If we choose to consider a U(1) theory where all
fermions have the same U(1) gauge charge, the total fermionic particle
number is constrained, in equilibrium, to be zero.  However this
restriction does not apply for nonabelian gauge theories, or for the
other conserved particle numbers.  There are therefore a large number of
fermionic number densities 
with diffusion constants we can consider.  In general
we should permit the diffusion ``constant'' to be a matrix in the flavor
space of conserved currents.  However, to
leading order in $\nf$ all the diffusion constants prove to be the same,
so we will ignore this complication.  In the U(1) theory, the diffusion
of the total fermionic number density is related to the behavior of infrared
electric fields and turns out to give the electrical conductivity; this
is discussed more in \cite{paper1}.

\subsection{Kinetic theory}

The stress tensor arises predominantly from fermionic degrees of
freedom, simply because there are $O(\nf)$ of them, and only $O(1)$
gauge degrees of freedom.  Naturally, fermionic currents and densities
are determined entirely by the fermionic degrees of freedom.
Furthermore, because $g^2 \sim 1/\nf$ is very small, the stress-energy
and fermionic currents are determined up to small correction by the
fermionic two point functions.
Hence it is sufficient, to determine transport coefficients to leading
order in $1/\nf$, to determine the departure from equilibrium of the
fermionic two point function.  This is precisely the purpose of kinetic
theory, so we turn to it next.

It is not our intention in this paper to derive kinetic theory; we feel
that the groundwork has been adequately established elsewhere
\cite{KadanoffBaym,CalzettaHu,Jeon}.  However, we will outline the
parametric argument that kinetic theory is applicable to the current
problem.  The easiest way to see that a kinetic theory description is
possible is to write down all bubble graphs
which appear at $O(\nf^1)$ and at
$O(\nf^0)$.  These are shown in Fig.~\ref{fig:bubbles}.  Cutting one
line in a graph gives the propagator to the order of interest.  The
leading order diagram is an undecorated fermion loop.  Cutting it gives
the bare fermionic propagator, which 
tells us that the fermionic propagator is at leading order a
free propagator.  (This is obvious since the lowest order fermionic
self-energy is $O(g^2\sim 1/\nf)$.)
Cutting the $O(\nf^0)$ graph on a gauge boson line shows
that the gauge boson propagator is at leading order a resummation of
one loop self-energy insertions, while cutting on a fermionic line shows
that the leading fermionic self-energy correction involves such a resummed
gauge boson line and a bare fermionic line.
(The gauge boson self-energy is large and requires 
resummation because $\nf$ different fermions can
run in the loop, and we must treat $g^2 \nf$ as $O(1)$.)

This means that it will be possible to use kinetic theory to determine
transport coefficients in this theory, with errors suppressed by
$1/\nf$, provided that we carry out a resummation of the gauge boson
self-energy wherever a gauge boson line appears in any process.  
It is incorrect and unnecessary to include gauge bosons as degrees of
freedom of the kinetic theory; incorrect because they do not have a
sharp mass shell, and unnecessary because there are so few gauge boson
degrees of freedom, so they have a negligible role both in carrying
conserved quantities and as out states in scattering processes.

The set of scattering processes needed in the kinetic
theory, neglecting $1/\nf$ corrections, is very small, and it will turn
out that the scattering processes are simple enough that they can be
treated without uncontrolled approximations.  Note that the large $\nf$
expansion is much more restrictive than the large $\nc$ expansion in
QCD; whereas at large $\nc$ all planar diagrams survive at leading
order, at large $\nf$ the structure even of the first subleading
diagrams is still very simple.  In particular it is only at $O(1/\nf)$
that diagrams containing nonabelian 3-point vertices first appear.  This
is why, at the order of interest, there is no difference between
treating QCD and QED.

Another requirement for a kinetic treatment to be useful is that it is
possible to expand the space dependence of the two point function in
gradients and drop terms with more than 1 gradient.  This is permissible
in determining transport coefficients because they by definition
describe the plasma's response to an arbitrarily slowly varying
disturbance.  In the case of shear viscosity, we consider an arbitrarily
slowly varying, divergenceless velocity field.  In the case of
fermionic particle number diffusion, we consider a slowly spatially
varying chemical potential $\mu$, which we take to be small%
\footnote{%
	Nothing prevents considering the case where the chemical
	potential is large, $\mu \gsim T$,
	provided $\beta \mu \ll \nf$ (to avoid complicated physics at
	the Fermi sphere).  However, for $\mu \gsim T$
	the transport coefficients must be reported
	as a function of $g^2\nf$ {\em and} 
	$\beta \mu$, and the energy density becomes $\mu$ dependent,
	so spatial gradients in $\mu$ also cause bulk flow.  We choose
	$\beta \mu \ll 1$ to avoid these complications.%
	}%
, $\beta \mu \ll 1$.
Slowly varying means that the gradients must vary on scales well larger
than $\beta \nf$.

In this regime we can write down a Boltzmann equation for the fermionic
population function (1 particle density matrix) $f^s(\p,\x,t)$ (with $s$
a spin, flavor, particle/anti-particle index), describing its time
evolution: 
\begin{equation}
    \left[
	{\partial \over \partial t}
	+
	\v_\p \cdot {\partial \over \partial \x}
	+
	{\bf F_{\rm ext}} \cdot {\partial \over \partial \p}
    \right]
    f^s(\p,\x,t)
    =
    -C[f] \, ,
\la {eq:Boltz}
\end{equation}
with $\v_\p = \hat\p$ the particle velocity,
$F_{\rm ext}$ an external force (if any), and $C$ the collision
integral, discussed more below.  Neither shear viscosity nor fermionic
number diffusion require including $F_{\rm ext}$, so we drop it
henceforth.  The collision integral is zero for the equilibrium
population function
\begin{equation}
f_0^s(\p) = \Big[ 1 + \exp ( -\beta_\mu p^\mu - q^s \beta \mu ) \Big]^{-1}
	\, ,
\end{equation}
where $q^s$ is the charge of species $s$ 
under the fermionic number under consideration.  The sign of $q^s$ is
opposite between particle and anti-particle.
We will expand in the small departure from equilibrium, which is
justified since
the size of the spatial scale of variation of $\beta_\mu(\x)$, $\mu(\x)$
is very large;%
\footnote{%
	The distinction between $f_0$ and $\delta f$ is not unique 
	without an additional prescription for $\delta f$ and
	$\beta_\mu$; we use the Landau-Lifshitz convention, under 
	which the sum over all particles' $\delta f$ 
	gives zero contribution to the 4-momentum and globally conserved
	charges.%
	}
\begin{equation}
f^s(\p,\x,t) = f_0(\p,\beta(\x),\mu(\x)) 
	+ f_0 (1{-}f_0)\; \delta f^s(\p,\x,t) \, .
\end{equation}
$\delta f^s$ is the quantity we are after, since the transport
coefficients are determined by it.  The nonequilibrium piece of the 
stress tensor is
\begin{equation}
T_{ij} - \delta_{ij} {\cal P} = \sum_{s} \int \frac{d^3 p}{(2\pi)^3}
	|\p| \: \hat p_i \hat p_j \, f_0(1{-}f_0) \; \delta f^s(\p) \, ,
\label{eq:Tij}
\end{equation}
and the current of species $s$ is
\begin{equation}
j_i^s = \sum \int \frac{d^3 p}{(2\pi)^3} \,
	q^s \, \hat p_i \, f_0(1{-}f_0) \; \delta f^s(\p) \, ,
\label{eq:j}
\end{equation}
with the sum over particle and anti-particle.
Hence, determining $\delta f$ in terms of $\partial_i u_j$ and $\nabla_i
\mu$ will determine the transport coefficients.

Since the gradient is taken small, $\delta f \ll f_0$ and we can
linearize in $\delta f$.  The spatial variation, {\em both} of $f_0$ and
$\delta f$, is slow; so we need only keep the gradient terms on the LHS
of Eq.~(\ref{eq:Boltz}) when they act on $f_0$.  The time derivative
term is also irrelevant \cite{paper1}.  The LHS of the 
Boltzmann equation becomes
\begin{equation}
\beta f_0(\p) [1{-}f_0(\p)] \left[ q^s \hat{p}_i \nabla_i \mu^s
	+ \frac{|\p|}{2} \left( \hat p_i \hat p_j - \frac{1}{3}
	\delta_{ij} \right) \left( \nabla_i u_j + \nabla_j u_i
	- \frac{2}{3} \delta_{ij} \naBla \cdot \u \right) \right]
	\, .
\label{eq:Boltz2}
\end{equation}
In writing the term involving $\nabla_i u_j$ in this form
we have used the restriction that we only consider divergenceless flow,
$\naBla \cdot \u = 0$.
The linearized collision operator is rotationally
invariant, so $\delta f$ will have the same angular dependence on $\hat
\p$ as the left hand side.  Since the two terms have different angular
dependence (the chemical potential involves an $\ell=1$ spherical
harmonic while the velocity gradient involves an $\ell=2$ spherical
harmonic), the two problems can be treated separately.
We will define a particle's ``charge'' $q$ to be $q^s$ for
the case of number diffusion and $|\p|$ for the case of shear
viscosity.  It is also convenient to introduce \cite{paper1}

\begin {equation}
    I_\ij(\hat \p)
    \equiv
    \cases
	{
	\;\hat p_i \,,\vphantom {\Big|} & \mbox{(diffusion)} \cr
	\sqrt {3\over2} \, (\hat p_i \hat p_j - {1\over3} \delta_{ij}) \,, 
	& \mbox{(shear viscosity)} \cr
	}
\label {eq:Iij}
\end {equation}

and

\begin {equation}
    X_\ij(x)
    \equiv
    \cases
        {	
	\;\nabla_i \, \mu \,, & \mbox{(diffusion)} \cr
	\frac{1}{\sqrt{6}} \left( \nabla_i u_j + \nabla_j u_i
	- {2\over3}\delta_{ij} \nabla \cdot \u \right) .
	& \mbox{(shear viscosity)} \cr
 	}
\label {eq:Tdrive}
\end {equation}
$X_\ij$ parameterizes the source for the departure from
equilibrium.  The LHS of Eq.~(\ref{eq:Boltz2}) is
proportional to $X_\ij(x) I_\ij(\hat\p)$.  The normalization is chosen
so that $I_\ij(\hat\p) I_\ij(\hat\p) = 1$.  For two general vectors $\p$
and $\k$,
\begin{equation}
I_\ij(\hat\p) I_\ij (\hat\k) = P_\ell ( \hat\p \cdot \hat\k ) \, ,
\end{equation}
with $P_\ell$ the $\ell$'th Legendre polynomial, and
with $\ell$ the spherical harmonic represented by $I_\ij$, which is the
same as the number of indices $I_\ij$ carries.

Rotational invariance of the collision integral ensures that we can
write the departure from equilibrium as
\begin{equation}
\delta f^s(\p,\x) = \beta^2 X_\ij(x) \chi^s_\ij(\p)
	\, , \qquad
	\chi^s_\ij(\p) = I_\ij(\hat\p) \chi^s(|\p|) \, .
\end{equation}
The function $\chi^s(|\p|)$ depends only on the magnitude of the
momentum and characterizes the departure from equilibrium of the
plasma.  It has the size of the source for departure from
equilibrium, $X_\ij$, and the angular dependence of the departure, $I_\ij$,
scaled out from it, and is the most convenient variable in terms of
which to write the Boltzmann equation.  Defining a ``normalized source''
for the departure from equilibrium,
\begin {equation}
    S^s_\ij(\p) \equiv
    -T \, q^s f_0(\p,x) [1 {-} f_0(\p,x) ] \>
    I_\ij(\hat \p) \,,
\end {equation}
(where again, for diffusion $q^s$ is defined above and 
for shear viscosity $q^s= |\p|$),
the linearized Boltzmann equation, \Eq{eq:Boltz2}, becomes
\begin{equation}
    S^s_\ij(\p)
    =
    \left(\C  \chi_\ij\right)^s(\p) \,.
\la {eq:lin1}
\end{equation}
The exact form of the collision operator $\C$ will be presented below.
The solution $\chi_\ij$ is simply related to the transport coefficients
\cite{paper1}.  Introducing the inner product (under which the collision
operator $\C $ is Hermitian and positive semidefinite)
\begin {equation}
\label{eq:innerprod}
    \Big( f,g \Big) \equiv \beta^3 \sum_a \int 
	\frac{d^3 p}{(2\pi)^3} \, f^a(\p) \, g^a(\p)
    \,,
\end {equation}
and considering the case of a fermionic number where all fermions have
$q^2 = 1$, using \Eq{eq:Tij} and \Eq{eq:j} and the definitions of the
transport coefficients leads to
\begin{eqnarray}
\eta & = & \frac{1}{15} \Big( \chi_{ij} , \C  \chi_{ij} \Big) \, ,
\nonumber \\
D & = & \frac{1}{\nf T^2} \Big( \chi_i , \C  \chi_i \Big) \, .
\label{eq:answer}
\end{eqnarray}
The coefficient in front of $D$ would be $1/3$, except that $D$ is
defined as $-j_i / \nabla_i j_0$, not $-j_i/\nabla_i \mu$.  The ratio
$j_0 / \mu$ is the charge susceptibility, and for small $\mu$ a simple
calculation gives $j_0/\mu = \nf T^2/3$.

\subsection{Variational method}

The departure from equilibrium $\chi(|\p|)$ is determined by
\Eq{eq:lin1}, which is an integral equation.  (This will become clear
when we write the form of the collision integral explicitly, \Eq{eq:C1}
below.) Integral equations are
difficult to solve unless the integration kernel $\C $ has a
particularly simple form, which is not the case for us.
Abandoning an exact solution, we can still get a
very accurate solution for $\chi$, and a more accurate determination of
$\eta$ and $D$, by expressing the problem as a variational one, writing
down a trial function for $\chi(|\p|)$ and
solving for a finite number of variational coefficients.  The accuracy
can then be improved, in principle without limit, by enlarging the basis
of variational functions considered.  This technique
constitutes a controlled approximation, meaning that it can
be made arbitrarily accurate.  It does not in practice limit the
accuracy with which we can extract transport coefficients.
This strategy has a very long
history in the solution of kinetic equations, see for instance
\cite{deGroot}.  Here we follow the approach of \cite{paper1}.  
\Eq{eq:lin1} is satisfied at the variational extremum of
\begin{equation}
Q(\chi) \equiv \Big( S_\ij \, , \, \chi_\ij \Big) - \frac{1}{2}
	\Big( \chi_\ij \, , \, \C  \chi_\ij \Big) \, ,
\label{eq:Q}
\end{equation}
and the value of the extremum is
\begin{equation}
Q_{\rm max} = \frac{1}{2} 
	\Big( \chi_\ij \, , \, \C  \chi_\ij \Big) \, .
\end{equation}
This extremal value is simply 
related to the transport coefficients, presented in
\Eq{eq:answer}.  Since the transport coefficients correspond to the
value of $Q$ at a maximum, the error in their determination is quadratic
in the error between a trial $\chi$ and the optimal $\chi$.  Hence,
extremizing over a suitably flexible \ansatz\ for $\chi$ will give a
very accurate value for the transport coefficients.  Following
\cite{paper1}, we consider the \ansatz
\begin{equation}
\chi^s(p) = \sum_{m=1}^N a_m q^s \phi^{(m)}(p) \, ,
\end{equation}
with $p = |\p|$ and $\phi^{(m)}$ some set of test functions.  
Our choice is
\begin{equation}
\phi^{(m)}(p) = \frac{(p/T)^m}{(1+p/T)^{N-1}} \, , \qquad 
	m=1,\ldots,N \, ,
\end{equation}
but other choices also work and lead to the same numerical answer.
Note that the $\phi^{(m)}$ for a small value of $N$ are each linear
combinations of the $\phi^{(m)}$ for a larger $N$, so increasing $N$
strictly increases the span of the \ansatz.  

Inserting this \ansatz\ for $\chi$ into \Eq{eq:Q} turns it into a
quadratic equation for the coefficients $a_m$;
\begin {equation}
    \tilde Q[\{a_m\}]
    =
    \sum_{m=1}^N
    a_m \, \tilde S_m
    -\half \sum_{m,n=1}^N
    a_m \, \tilde C_{mn} \, a_n \,,
\end {equation}
with $\tilde S_m$ given by
$
    \tilde S_m \equiv \Big(S_\ij, q \phi_\ij^{(m)} \Big)
$,
and similarly
$
    \tilde C_{mn} \equiv \Big(q \phi_\ij^{(m)}, 
	\, \C q \phi_\ij^{(n)} \Big)
$, where $\phi^{(m)}_\ij(\p) \equiv I_\ij(\hat\p) \phi^{(m)}(p)$. 
Maximizing $\tilde Q$ is now a trivial linear algebra exercise
which gives
$
    a = \tilde C^{-1} \, \tilde S
$
and
\begin {equation}
Q_{\rm max} = \frac{1}{2} \tilde S^\trans \, \tilde C^{-1} \, \tilde S
    \,,
\label {eq:sigmaa}
\end {equation}
where $a = ||a_m||$ and $\tilde S = ||\tilde S_m||$
are the $N$-component coefficient and source vectors, respectively,
in the chosen basis, and $\tilde C \equiv || \tilde C_{mn} ||$
is the (truncated) collision matrix.
The determination of transport coefficients then hinges on the accurate
evaluations of the integrals
$\Big(S_\ij, q \phi_\ij^{(m)} \Big)$ and
$\Big(q \phi_\ij^{(m)}, \, \C q \phi_\ij^{(n)} \Big)$.
The integral involved in determining 
$\tilde S_m$ is 1 dimensional and is easy to evaluate, quickly and
accurately, by numerical quadratures; we do not discuss it further.
Evaluating $\tilde C_{mn}$ is the topic of the next section.
However it is already possible to determine the power of $\nf$ which
will appear in the final answer.
The quantity $\tilde S$ contains a sum over all species but no powers of
the coupling constant, so it is $O(\nf)$.  The collision integral, since
it arises from an $O(1)$ bubble diagram, must be $\tilde C_{mn} \sim
\nf^0$.  The appropriate powers of $T$ follow easily on dimensional
grounds.  Therefore $\eta \propto \nf^2 T^3$, and $D \propto \nf/T$
(because there is an explicit negative power of $\nf$ in
\Eq{eq:answer}).

\section{Collision integral}
\label{sec:collision}

\subsection{Relevant diagrams}

\begin{figure}
\centerline{\epsfxsize=4in\epsfbox{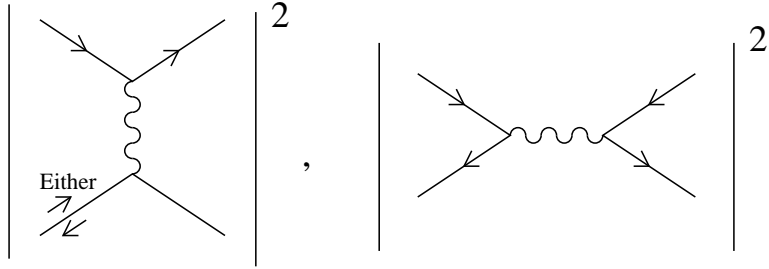}}
\caption{\label{fig:scatter} Scattering processes which must be
considered at leading order .}
\end{figure}

To make more progress we must analyze what collision processes are
important in this theory.  They are given by cutting the fermionic
self-energy diagram of Fig.~\ref{fig:bubbles}, see
Fig.~\ref{fig:scatter}.  Two scattering processes are important, $t$
channel scattering from a particle or anti-particle and $s$ channel
annihilation and creation of a new pair.  The external lines in these
processes may all be treated as massless, on-shell particles, with
uncorrected dispersion relations, since the self-energy of the fermion
is $O(g^2 \sim 1/\nf)$.  This is not so for the gauge boson propagator
appearing in the diagram, which must be treated with care.

\begin{figure}
\centerline{\epsfxsize=4.8in\epsfbox{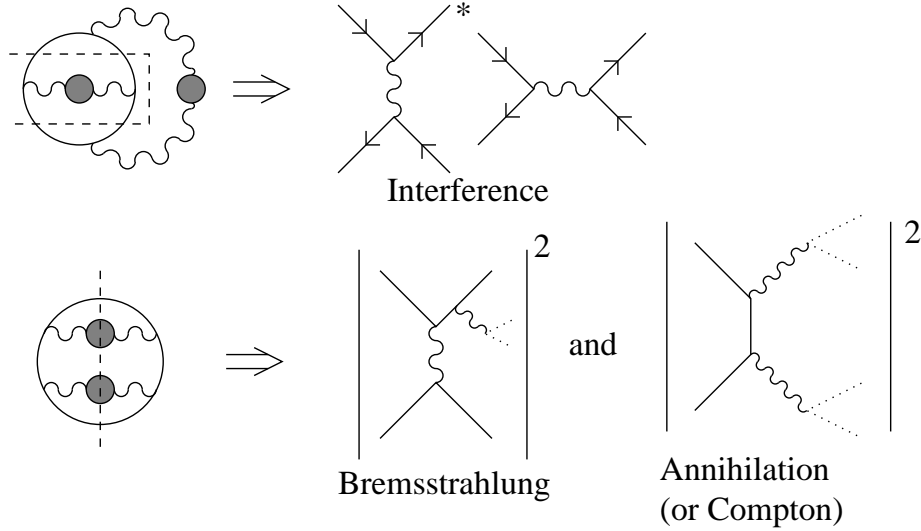}}
\caption{\label{fig:absent} Scattering processes which may be important
at leading order in the coupling $g$, when we do not expand in $\nf$,
but which are $\nf$ suppressed and can be neglected here.  Dashed lines
on the bubble graphs show where they are cut to give the diagrams
presented.  Dotted lines in the scattering processes indicate that gauge bosons
are not propagating states in the kinetic theory, and we must
consider the process including the states a gauge boson eventually
scatters against or decays into.}
\end{figure}

Note that only $2 \leftrightarrow 2$ processes appear in the collision
integral to the order of interest, and that the external lines are all
fermions.  The set of diagrams required is much simpler than the set
which are needed 
at leading order in $g$ when not adopting the large $\nf$ expansion.  In
that case, Compton scattering and annihilation to gauge bosons
are already present at leading logarithmic order \cite{paper1}, and at
leading order interference diagrams between external leg assignments of
the $s$ and $t$ channel processes are also present.  Similarly, in the
current context, bremsstrahlung emission is also suppressed.  These
``notable for their absence'' diagrams are summarized in
Fig.~\ref{fig:absent}.  In each case there is a clear power counting
reason for the diagram's absence.  Interference between external legs is
only possible when all external lines are of the same
species; but this only occurs in $1/\nf$ of scattering events.
Similarly, Compton scattering or annihilation to gauge bosons involve a sum
over only one, not two, fermionic species indices\footnote
	{%
	In fact, since only the fermions may be treated with kinetic 
	theory, a gauge boson is never allowed as an outgoing particle; 
	instead we must draw the eventual scattering the gauge boson
	undergoes and include those fermions as external lines.  Since
	the gauge boson inverse propagator has an 
	$O(g^4 \nf^2 \sim 1)$ imaginary part where the real part
	vanishes, the integration over the gauge boson ``mass shell'' is not
	dangerous.  Drawing the diagrams including the fermionic lines 
	makes clearer the absence of Compton and
	bremsstrahlung processes in the current analysis; they
	really represent either $3 \leftrightarrow 3$ or $2
	\leftrightarrow 4$ processes.  If we were considering
	bulk viscosity these processes would be important.%
	}.
Another way to see the suppression of these processes is to consider
what bubble graph must be cut to produce them; this is also shown in
Fig.~\ref{fig:absent}, and in each case the bubble diagram is
$O(\nf^{-1})$.  

\subsection{Form of $\tilde C_{mn}$}

Since only $2 \leftrightarrow 2$ processes are important, the collision
integral can be written \cite{paper1}
\begin{eqnarray}
\label{eq:C1}
(\C  \chi_\ij)^a(\p)  & = & \frac{1}{2} \sum_{bcd}
	\int \frac{d^3 \k d^3 \p' d^3 \k'}{(2\pi)^9 16 pkp'k'}
	(2\pi)^4 \delta^3(\p+\k-\p'-\k') \delta(p+k-p'-k')
	\nonumber \\ & & \qquad \quad \times
	\left| {\cal M}^{ab}_{cd}(\p,\k,\p',\k') \right|^2
	f_0^a(p) \, f_0^b(k)[1{-}f_0^c(p')][1{-}f_0^d(k')]
	\nonumber \\ & & \qquad \quad \times
	\Bigl[
	    \chi^a_\ij(\p) + \chi^b_\ij(\k) -
	    \chi^c_\ij(\p') - \chi^d_\ij(\k')
	\Bigr] \,.
\end{eqnarray}
Our notation is that the incoming momenta are $\p$ and $\k$ and the
outgoing momenta are $\p'$ and $\k'$; and as before, $k = |\k|$, and so
forth.  ${\cal M}^{ab}_{cd}(\p,\k,\p',\k')$ is the matrix element (in
relativistic normalization) for $(ab)$ to go to $(cd)$.  We have already
enforced on-shell delta functions on the external 4-momenta using free
massless dispersion relations, so the energies of the external particles
equal the magnitudes of their momenta.  
The leading factor of $(1/2)$
corrects for the $c \leftrightarrow d$ double counting of the sum, or
for the outgoing symmetry factor when $c=d$.  

The quantity we want is $\tilde{C}_{mn}$, that is, \Eq{eq:C1} with
$\chi_{ij}^s$ replaced with $q^s \phi^{(m)}_\ij$, 
multiplied by
$q^a \phi^{(n)}_\ij(\p)$, integrated over $\p$, and summed over $a$.  The
resulting phase space integral 
and matrix element are symmetric on interchange of
the $(abcd)$ indices, so we can also symmetrize the $\phi$ factors,
yielding 
\begin{eqnarray}
\label{eq:Cmn}
\Big( q \phi^{(m)}_\ij \, , \, 
	\C q \phi^{(n)}_\ij \Big) & = & 
	\frac{\beta^3}{8} \sum_{abcd}
	\int \frac{d^3 \p d^3 \k d^3 \p' d^3 \k'}{(2\pi)^{12} 16 pkp'k'}
	(2\pi)^4 \delta^3(\p+\k-\p'-\k') \delta(p+k-p'-k')
	\nonumber \\ & & \qquad \quad \times
	\left| {\cal M}^{ab}_{cd}(\p,\k,\p',\k') \right|^2
	f_0^a(p) \, f_0^b(k)[1{-}f_0^c(p')][1{-}f_0^d(k')]
	\nonumber \\ & & \qquad \quad \times
	\Bigl[
	    q^a \phi^{(m)}_\ij(\p) + q^b\phi^{(m)}_\ij(\k) -
	    q^c\phi^{(m)}_\ij(\p') - q^d\phi^{(m)}_\ij(\k')
	\Bigr]
	\nonumber \\ & & \qquad \quad \times
	\Bigl[
	    q^a \phi^{(n)}_\ij(\p) + q^b\phi^{(n)}_\ij(\k) -
	    q^c\phi^{(n)}_\ij(\p') - q^d\phi^{(n)}_\ij(\k')
	\Bigr] \, .
\end{eqnarray}
It is clear from this expression that $\C$ is positive semidefinite.
For $\ell \neq 0$, it is in fact positive definite.\footnote
	{It is not positive definite for $\ell=0$, in which case 
	the collision integral vanishes if $\phi(p)$ is a
	constant, until we include higher order, number changing
	processes.  This is discussed more in \pcite{Jeon}, and is
	the reason that evaluating bulk viscosity is difficult.}

The matrix element gets 3 contributions; the $s$ channel contribution,
the $t$ channel contribution for scattering of like sign
(both particle or both anti-particle) fermions, and the $t$ channel
contribution for scattering of opposite sign fermions (a particle and an
anti-particle).  The matrix element for the $s$ channel process, summed
over all species indices $(abcd)$ and including the symmetry factor of
$(1/8)$, is
\begin{equation}
\label{eq:schann}
{\rm s \; channel: } \quad
\frac{1}{8} \sum_{abcd} \left| {\cal M}^{ab}_{cd} \right|^2
= 4 (g^2 \nf)^2 \ttuuss \, ,
\end{equation}
for the U(1) theory.  We discuss the nonabelian generalization in the
next section.
The meaning of ``[$t^2+u^2$$/s^2$]'' is a quantity which equals this
(with $s,t,u$ the usual Mandelstam variables) in
the small $g^2 \nf$ limit, and is given in complete detail in Appendix
\ref{app:Msq} \Eq{eq:s_M}.  The expression for $t$ channel, like sign
scattering is
\begin{equation}
\label{eq:tchann}
{\rm t \; channel \; like \; sign: } \quad
\frac{1}{8} \sum_{abcd} \left| {\cal M}^{ab}_{cd} \right|^2
= 4 (g^2 \nf)^2 \ssuutt \, .
\end{equation}
The expression for opposite sign scattering is identical.  Again, the
quantity ``[$s^2 + u^2$$/t^2$]'' equals this in the small $g^2 \nf$
limit, but is given by a more complicated expression presented in
Appendix \ref{app:Msq}, \Eq{eq:t_M}.

\subsection{Doing the integrals}

To compute transport coefficients it remains to perform the integrals
presented in \Eq{eq:Cmn}.  There is a 12 dimensional integral to
perform.  Four of the integrals will be performed by the energy-momentum
conserving delta functions, and three are trivial because the way
we structured \Eq{eq:Boltz2} makes \Eq{eq:Cmn} invariant under global
rotations.  This leaves 5 integrals.  Appendix \ref{app:phasespace}
shows two nice ways of parametrizing these remaining integrals.  One way,
due to \cite{BMPRa}, is convenient for $t$ channel exchange, and another
is convenient for $s$ channel exchange.

We can write \Eq{eq:Cmn} as a sum of two 5 dimensional integrals, one
containing the $s$ channel contribution to $|{\cal M}|^2$ and the other
containing the $t$ channel contributions, by using
\Eq{eq:int_t} and \Eq{eq:int_s}.  The matrix elements are given by
\Eq{eq:schann} with \Eq{eq:s_M}, and twice \Eq{eq:tchann} with \Eq{eq:t_M}.
The product of $\phi$ functions is evaluated for shear viscosity using
\begin{equation}
|\p| \phi^{(m)}_{ij} ( \hat\p ) |\k| \phi^{(n)}_{ij} ( \hat\k )
	= pk \phi^{(m)}(p) \phi^{(n)}(k) P_2 ( \cos \theta_{pk} ) \, ,
\end{equation}
and similarly for the other terms when the product of $\phi$'s is
expanded.  Here $P_2(x)=(3x^2-1)/2$ is the second
Legendre polynomial, and all the cosines of angles are given in
Appendix \ref{app:phasespace}.  For both the $s$ and $t$ channel part of
the scattering integral, there is an integration over an azimuthal angle
$\phi$ which can be done analytically, leaving a 4 dimensional integral,
over the frequency $\omega$ and momentum $q$ carried by the gauge boson line
and two external momenta.  The integrand is long and unenlightening and
can be obtained from the equations listed above.  We perform it by
numerical quadratures using adaptive mesh refinement
methods.  The accuracy with which the integrals can be performed limits
the accuracy of the determination of the transport coefficients, but it
is not too difficult to achieve $0.1\%$ accuracy in the finally determined
transport coefficients, for the full range of the coupling constant $g^2
\nf$ that we consider.  Note that, besides the explicit $g^2 \nf$ dependence
of \Eq{eq:schann} and \Eq{eq:tchann}, there is also
quite complicated $g^2 \nf$ dependence in the matrix elements because of
the gauge boson self-energies.  Therefore the full integration must be
re-performed for every value of $g^2 \nf$ under study.  Note also that
$g^2 \nf$ requires (vacuum) renormalization.  We regularize
using the \MSbar\ scheme.  We also choose the
renormalization point\footnote{Do not confuse the \MSbar\
renormalization point $\bar\mu$ with the
chemical potential $\mu$.}  $\bar\mu$ to be the ``dimensional
reduction'' value \cite{KLRS} $\bar{\mu}_{\rm DR} = \pi T
\exp(-\gamma_{\rm E})$, with $\gamma_{\rm E} = .577\ldots$ the
Euler-Mascheroni constant.  This choice is arbitrary and
represents a choice in the presentation of the results only.  Its
motivation is that, at this renormalization point only, the free energy
of an infrared electromagnetic field is given by $\int d^3x (\E^2+\B^2)/2$.

We also cannot treat arbitrarily large 
$g^2 \nf$ because the presence of the Landau pole in the
theory becomes problematic.  Requiring that the Landau pole
occur further into the ultraviolet than
$|q^\mu q_\mu| = (40 T)^2$ requires $g \sqrt{\nf} \leq 4.4$.  
A Landau pole this far in the ultraviolet does not affect our 
calculation because the contribution to \Eq{eq:Cmn} from integration
regions with $|\omega^2 - q^2|$ large enough to encounter the Landau pole
is exponentially small.  In fact such regions were never sampled by
the adaptive integration routine.  For $g \nf^{1/2} = 3$, the
energy scale where the pole appears is already above $1000T$.

For the case of fermion number diffusion, there are some additional
cancellations which make the structure of the integrand slightly
simpler than for shear viscosity.  Unlike shear viscosity, the charge
$q^s$ for number diffusion is opposite for a particle and its
anti-particle.  Further, the $t$ channel matrix element to scatter from a
particle equals the matrix element to scatter from its anti-particle (up
to $1/\nf$ suppressed corrections).
Hence, for $t$ channel exchange, summing on like and
opposite sign scatterings cancels all ``cross-terms'' 
in \Eq{eq:Cmn} with both $p$
and $k$ line arguments for $\phi$'s.  Using also the $\p \leftrightarrow
\k$ symmetry of the integration, and the fact that the incoming and
outgoing particles $a$ and $c$ are the same type and have the same
charge, one may substitute
\begin{eqnarray}
&& \sum_{abcd} |{\cal M}|^2(t{\rm \; channel})
	\Bigl[
	    q^a \phi^{(m)}_i(\p) + q^b\phi^{(m)}_i(\k) -
	    q^c\phi^{(m)}_i(\p') - q^d\phi^{(m)}_i(\k')
	\Bigr]
	\nonumber \\ & & \hspace{1.3in} \times
	\Bigl[
	    q^a \phi^{(n)}_i(\p) + q^b\phi^{(n)}_i(\k) -
	    q^c\phi^{(n)}_i(\p') - q^d\phi^{(n)}_i(\k')
	\Bigr] 
	\nonumber \\
& \Rightarrow & \sum_{abcd} 2|{\cal M}|^2(t{\rm \; channel})
	(q^a)^2 \Bigl[
	    \phi^{(m)}_i(\p) - \phi^{(m)}_i(\p') \Bigr]
	\Bigl[
	    \phi^{(n)}_i(\p) - \phi^{(n)}_i(\p') \Bigr] \, .
\end{eqnarray}
For the $s$ channel exchange a
similar simplification occurs because in that case $q^c = -q^d$ but the
integral is symmetric on $\p' \leftrightarrow \k'$ (as well as on
$(\p,\k) \leftrightarrow (\p',\k')$); so 
\begin{eqnarray}
&& \sum_{abcd} |{\cal M}|^2(s{\rm \; channel})
	\Bigl[
	    q^a \phi^{(m)}_i(\p) + q^b\phi^{(m)}_i(\k) -
	    q^c\phi^{(m)}_i(\p') - q^d\phi^{(m)}_i(\k')
	\Bigr]
	\nonumber \\ & & \hspace{1.3in} \times
	\Bigl[
	    q^a \phi^{(n)}_i(\p) + q^b\phi^{(n)}_i(\k) -
	    q^c\phi^{(n)}_i(\p') - q^d\phi^{(n)}_i(\k')
	\Bigr] 
	\nonumber \\
& \Rightarrow & \sum_{abcd} 2|{\cal M}|^2(s{\rm \; channel})
	(q^a)^2 \Bigl[
	    \phi^{(m)}_i(\p) - \phi^{(m)}_i(\k) \Bigr]
	\Bigl[
	    \phi^{(n)}_i(\p) - \phi^{(n)}_i(\k) \Bigr] \, .
\end{eqnarray}
(The same
symmetries were used in \cite{paper1} to allow a simple leading log
calculation of the diffusion constant.)  For both the $s$ and $t$
channel integrations, there is an azimuthal integration which is trival,
and the two external momentum integrations factor and can be performed
separately (rather than being nested).  This simplifies somewhat the
numerical integrations.

Further, the
cancellations just described show that, for
number diffusion, the way that the departure from equilibrium in one
species evolves decouples from the departure from equilibrium in every
other species.\footnote{Note that this would not happen if we considered
chemical potentials 
$\mu \gsim T$.}  This is why all the fermionic diffusion constants are
the same, unless the total fermionic number is coupled to a gauged U(1)
field.  

\section{All-orders Results}
\label{sec:results}

\begin{figure}[p]
\centerline{\epsfxsize=5.5in\epsfbox{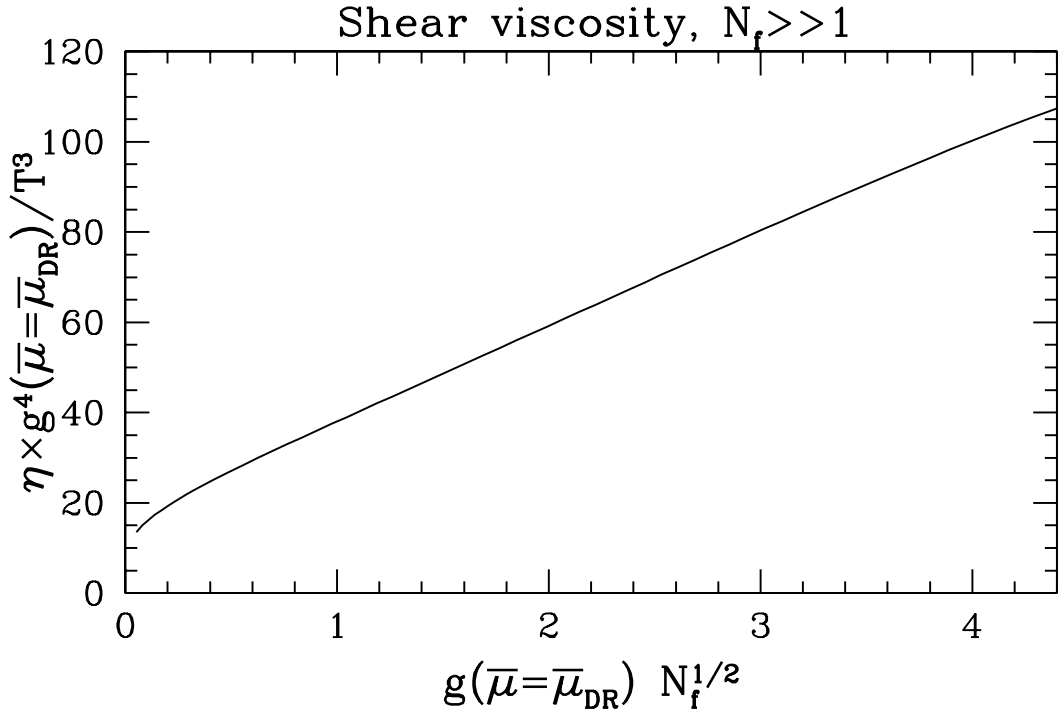}}
\vspace{0.1in}
\caption{Shear viscosity, computed
to all orders in $g^2 \nf$.  The renormalization point $\bar\mu$ in
the \MSbar\ scheme is $\bar\mu=\pi e^{-\gamma_{\rm E}} T$,
the ``dimensional reduction'' answer. \label{fig:result1}}
\end{figure}

\begin{figure}[p]
\centerline{\epsfxsize=5.5in\epsfbox{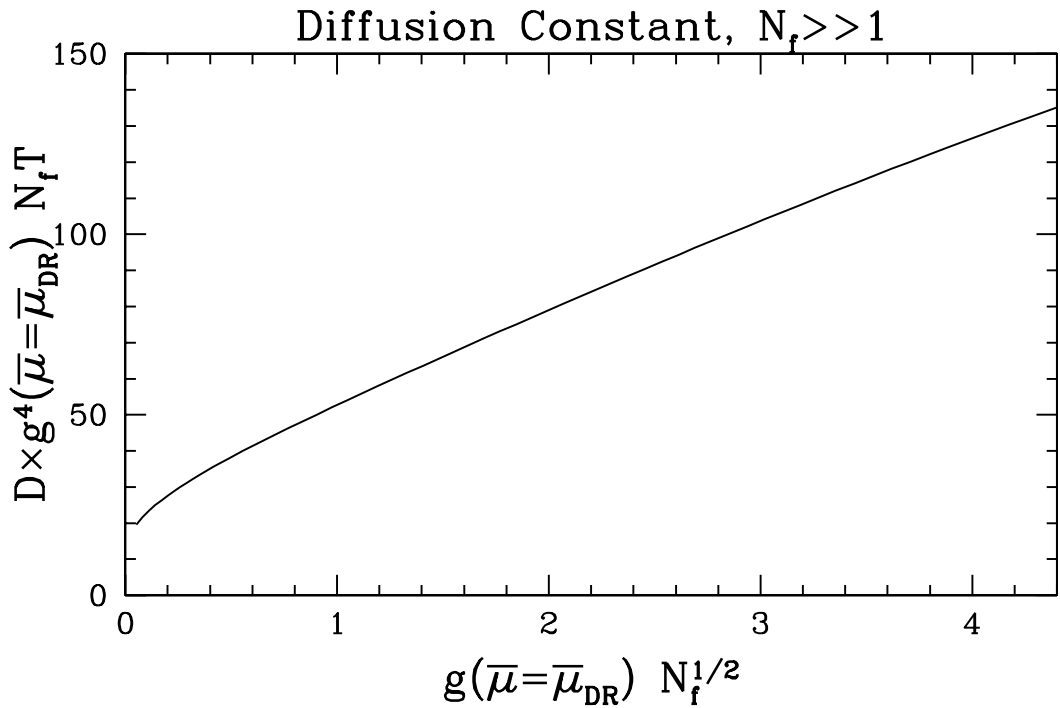}}
\vspace{0.1in}
\caption{Diffusion constant, computed
to all orders in $g^2 \nf$.  The renormalization point is the same 
as in Fig.~\ref{fig:result1}. \label{fig:result2}}
\end{figure}

The last two sections describe how the transport coefficients, at leading
order in $1/\nf$ but all orders in $g^2 \nf$, may be computed for $U(1)$
gauge theory.  The results appear in Figures \ref{fig:result1} and
\ref{fig:result2}.  
Here we discuss the results and generalize them to QCD,
with $\nc \ll \nf$ so as not to spoil the expansion we have used.

First we remark that, while there is no ``diffusion constant'' for the
{\em total\,} fermionic number density 
when it is coupled to a U(1) gauge field,
the would-be diffusion constant instead determines the  
electric conductivity for the U(1) gauge field
\cite{paper1}:
\begin{equation}
\sigma = \frac{g^2 \nf T^2}{3} \, D \, .
\label{eq:sigma}
\end{equation}
Therefore the results in Fig.~\ref{fig:result2} also constitute a result for
the electrical conductivity of this theory.  For this interpretation the
choice of renormalization constant is essential, since only for $\bar\mu
= \bar\mu_{\rm DR}$ does the free energy density stored in an electric
field equal $E^2/2$.

In presenting the transport coefficients in Figs \ref{fig:result1} and 
\ref{fig:result2}, we
factor out the leading $g^{-4}$ behavior from the
transport coefficients; but there remains nontrivial $g^2 \nf$ dependence.
This is because the size of the gauge boson self-energy, relative to the
tree gauge propagator, depends on $g^2 \nf$.  It is not surprising that
this is important for $g^2 \nf \gsim 1$.  In fact the dependence
continues down to arbitrarily small $g^2 \nf$, where the transport
coefficient scales as $1/\ln(1/g^2 \nf)$.  This is because the
$t$ channel contribution to the scattering integral, \Eq{eq:Cmn}, has a
logarithmic small $q$ divergence which is cut off by the presence of a
self-energy on the gauge boson line.
This is discussed further in the next section.

The results presented in Fig.~\ref{fig:result1} and
Fig.~\ref{fig:result2} are for U(1) gauge theory.  Consider
instead a general nonabelian theory, with $\nf$ flavors of 
fermions in a common
representation of the group, of dimension $\df$ and with quadratic
Casimir $\cf$, and trace normalization $\tf = \cf \df / \da$, with $\da$
the dimension of the adjoint representation.  For SU($\nc$) gauge
theory with fundamental representation fermions, $\cf = (\nc^2-1)/2\nc$,
$\df=\nc$, and $\tf = 1/2$.  The total number of
fermionic degrees of freedom is then $4 \df \nf$, rather than $4 \nf$ in
the U(1) theory.  (The 4 is because we consider Dirac fermions; there
are two spins, and particle/anti-particle.)  Therefore, 
the sum in \Eq{eq:innerprod} includes $4 \nf \df$ terms,
rather than $4 \nf$ terms as in the U(1) theory, so $\tilde S_m$ is
multiplied by $\df$.  Meanwhile, the gauge field self-energy has the
replacement 
\begin{equation}
\mD^2 = \frac{g^2 \nf T^2}{3} \quad \Rightarrow \quad
\mD^2 = \frac{g^2 (\tf \nf + \ta) T^2}{3}
	\simeq \frac{g^2 \tf \nf T^2}{3} \, .
\label{eq:mDQCD}
\end{equation}
The expression with $\ta$ ($\ta=\ca$ the adjoint Casimir, 
$\ta=\nc$ in SU($\nc$) gauge
theory) holds generally to leading order in $g^2$ in QCD, with the
$\ta$ term arising from gauge boson loops; in the last
expression we have enforced the large $\nf$ limit to eliminate this
nonabelian term.  The group theory coefficient for 
$\tilde C_{mn}$ is $(\df^2 \cf^2/\da)$.  Also, the fermion number
susceptibility, which goes into the determination of $D$, is multiplied
by $\df$; hence $D$ is smaller by this factor.
Therefore, to use the figures
to get the transport coefficients for a nonabelian theory, one should
interpret the $x$ axis in Fig.~\ref{fig:result1} as 
\begin{equation}
g \nf^{1/2} \quad \Rightarrow \quad g (\tf \nf)^{1/2} \, ,
\end{equation}
and replace the $y$ axis label with
\begin{equation}
\eta \times \frac{g^4}{T^3} \quad \Rightarrow \quad 
\eta \times \frac{g^4 \cf^2}{\da T^3} \, , \qquad
D \times g^4 \nf T \quad \Rightarrow \quad 
D \times \frac{g^4 \nf \cf^2 \df T}{\da} \, . 
\end{equation}
In SU($\nc$) QCD with fundamental fermions, $\tf=1/2$, $\df=\nc$, and
$\cf^2/\da = (\nc^2-1)/(4\nc^2)$.  In converting $D$ into $\sigma$ using
\Eq{eq:sigma}, the $g^2$ appearing in that equation should be the U(1)
gauge coupling, and there is an additional factor of $\df$,
\begin{equation}
\sigma =\frac{e^2 \df \nf T^2}{3} D \, ,
\end{equation}
with $e$ the fermionic coupling to the U(1) gauge field.

\section{Leading Order Calculation}
\label{sec:leading}

One use of our results is to allow a comparison against a leading order
calculation which relies on the hard thermal loop expansion.
To date, only leading logarithmic results are available.
Repeating the analysis of \cite{paper1} in this context gives
at leading logarithmic order
\begin{equation}
\label{eq:LL}
D = \frac{91.1725}{\nf g^4 T} \: \times \: \frac{1}{\ln(T/\mD) + O(1)} \, ,
\qquad
\eta = \frac{61.0727 T^3}{g^4} \: \times \: \frac{1}{\ln(T/\mD) + O(1)} \, .
\end{equation}
To study better the reliability of the HTL expansion, we present
here a leading order calculation.  We perform it in the spirit of other
leading order HTL calculations, which is to say, all self-energies are
replaced with just the thermal part of the self-energy in the HTL
approximation, wherever that treatment is parametrically justified.
Note that the HTL vertices all vanish at leading order in $1/\nf$ and
are not needed here.  In this section and this section alone, we
will treat $g^2 \nf$ as a small quantity which can be expanded in
parametrically.  Equivalently we treat the Debye mass, given by
$\mD^2 = g^2 \nf T^2 / 3$, as $\mD \ll T$.

It turns out that {\em both} the $t$ and $s$ channel exchange parts of
the collision integral depend in a nontrivial way on the presence of a
self-energy on the gauge boson line.  We will discuss each in turn.

\subsection{Leading order $t$ channel exchange}

It has long been appreciated that there is a subtlety in the $t$ channel
exchange process, and that inclusion of a self-energy on the gauge boson
lines is necessary, not only to get the correct leading order behavior,
but to get a behavior which does not suffer from infrared divergences in
the evaluation of the collision integral \cite{BMPRa}.  This is because
of the small denominator provided by the propagator when $\p-\p'$ is
small.  As shown in Appendix \ref{app:phasespace}, the $t$ channel phase
space integrals can be written \cite{BMPRa} (\Eq{eq:int_t}) 
\begin{equation}
    \frac{\beta^3}{2^{9} \pi^6}
    \int_0^{\infty} dq
    \int_{-q}^q d\omega 
    \int_{\frac{q-\omega}{2}}^\infty dp 
    \int_{\frac{q+\omega}{2}}^\infty dk
    \int_0^{2\pi} d\phi \, ,
\end{equation}
with $p$ and $k$ the energies of the two incoming particles, and
$\omega$ and $q$ the energy and momentum carried by the exchanged gauge
boson.  When the
particle energies are $O(T)$ but the exchange momentum is 
$\mD \ll q \ll T$, we can expand in small $q/p$ and simultaneously drop
the self-energy effects on the propagators, so that \Eq{eq:t_M} becomes
\begin{equation}
\ssuutt \simeq \frac{s^2+u^2}{t^2}
	\simeq \frac{8p^2 k^2 (1-\cos \theta)^2}{q^4} \, .
\end{equation}
At first sight the integral over $q$ is severely infrared divergent.
However, two more powers of $q$ arise from cancellations in the
expressions of the form $\Big[ \phi_\ij(\p) - \phi_\ij(\p') \ldots \Big]$ in
\Eq{eq:Cmn} \cite{paper1}, because $\p{-}\p' = \q$ is $O(q)$ 
and $\phi_\ij$ is a smooth function of its argument; so the
$\phi_\ij(\p)$ and $\phi_\ij(\p')$ terms cancel up to an 
$O(q)$ remainder.  Therefore, the $q$ integration is really of the form
\begin{equation}
\sim \int_{\sim \mD}^{\sim T} dq \int_{-q}^{q} d\omega \frac{1}{q^2} \, ,
\end{equation}
which is logarithmically sensitive to the $gT$ scale.  
The upper end of the logarithm is cut
off because the expansion $q \ll p$ is no good there.  The lower end of
the integration range is only cut off by the inclusion of the
self-energies in the scattering processes.  Therefore it is necessary to
include those self-energies to get a finite answer.  However, for $\mD
\ll T$, they only play a role when $q \lsim \mD \ll T$, which is clear
from the form of the matrix element, given in \Eq{eq:t_M}.

At leading logarithmic order it is only necessary to determine the {\em
coefficient} of the log divergence arising from the $\mD \ll q \ll T$
momentum range.  Such a
calculation was carried out in \cite{paper1}, where it was also not
necessary to appeal to a large $\nf$ expansion.  However, a
leading order calculation requires a treatment of both the $q \sim \mD$
and the $q \sim T$ momentum ranges, as well as the intermediate range.
For such a leading order calculation we must perform the full
integration presented in \Eq{eq:Cmn}.  However we may simplify the
treatment by replacing the full self-energy on the gauge boson line with
its HTL counterpart.  That is, we may {\em drop} the vacuum contribution
to the self-energies, and make the simplifying approximations for the
thermal parts which are presented in \Eq{eq:HTL}.  We may do this {\em
throughout} the integration range of \Eq{eq:Cmn}.  The vacuum part is
uniformly suppressed by $g^2 \nf$ (up to logarithms).  At small transfer
momenta $q \sim \mD$, the corrections to the HTL approximation for the
thermal contribution to self-energies is $O(\mD^2/T^2 \sim g^2 \nf)$,
and the approximation is valid.  At large momenta $q \sim T$, 
the HTL approximation is invalid, but the whole thermal self-energy is
an $O(g^2 \nf)$ correction and is negligible anyway.

Our method for achieving a leading order determination of the integrals
in \Eq{eq:Cmn} is therefore to replace the self-energy on the gauge
boson line with the HTL self-energy, over the full momentum integration
range.  The integral is then performed without further approximation
(such as expansion in $q \ll p$).

We have also tested the sensitivity to  a 
slight modification of the matrix element,
moving the $\cos \phi$ independent piece in \Eq{eq:t_M} which is
multiplied by $(q^2 + \tilde \PiT)^{-2}$ over to the piece multiplied by 
$(q^2 + \tilde \PiL)^{-2}$, so appearances of the transverse self-energy
coincide
with powers of $\cos \phi$.  This modification makes an $O(g^2 \nf)$
change because the piece moved is $O(g^2 \nf)$ suppressed where the
difference in self-energies matters.  It coincides with the separation
between transverse and longitudinal self-energies presented in
\cite{Heiselberg}, and is the easiest thing to implement, especially if
scatterings with gauge bosons are important (in a nonabelian theory and
not at large $\nf$).  We find the difference is relatively small,
comparable to the difference between the leading order at the
dimensional reduction renormalization point and the all
orders answer.  This is not the procedure we used to get the results
presented in the figures to follow.

\subsection{Leading order $s$ channel exchange}
\label{sec:schann}

Naively, the $s$ channel exchange contribution to \Eq{eq:Cmn} should not
need any self-energy insertion at all on the gauge boson propagator.
After all, the exchange energy is only small when both incoming fermions
have small momentum, which is a strongly phase space suppressed region.
Since we are working with fermions there is also no enhancement from the
statistical factors in this region.  However, while the self-energy is in fact
negligible at {\em generic} momenta, this is not so when the incoming
fermions are hard but {\em collinear}, $p,k \sim T$ but 
$1 {-} \hat \p \cdot \hat \k \sim g^2 \nf$, 
or equivalently 
$\omega - q \sim g^2 \nf T$.  Furthermore, this region turns out to
contribute $O(1)$ of the contribution of the entire $s$ channel
piece of \Eq{eq:Cmn}.  The reason is that the thermal self-energy
lifts the transverse gauge boson mass shell from lying on the light cone
$\omega^2 - q^2 = 0$ to lying in the timelike region, 
$\omega^2 - q^2 \sim \mD^2$.
Since the fermions still propagate at the speed of light, it is
kinematically allowed for a fermion pair to annihilate to an on-shell
gauge boson, which propagates for some time and then breaks up back into
fermions; that is, there is resonant enhancement of the scattering cross
section on the gauge boson ``mass shell''.  The phase space suppression
is compensated for by the on-shell near singularity from the propagator.

\begin{figure}[t]
\centerline{\epsfxsize=1.5in\epsfbox{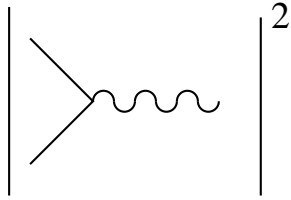}}
\caption{Diagram for production of an on-shell gauge
boson.\label{fig:2to1}}
\end{figure}

At finite $g^2 \nf$, this does not lead to any divergence in the
the matrix element at any value of $\omega \neq q$, 
because the gauge boson self-energy has a nonzero imaginary part
everywhere except the light cone.  
The imaginary part ``smears out'' the gauge boson mass
shell from a delta function into a sharply peaked Lorentzian.  We now
estimate the matrix element near resonance.  The
relevant part of the matrix element, \Eq{eq:schann}, is the term
involving the square of the transverse propagator, which is of the form 
(see \Eq{eq:s_M})
\begin{equation}
|{\cal M}|^2(s {\rm \; channel}) 
\sim \frac{(\omega^2 - q^2)^2}{|\omega^2 - q^2 - \PiT|^2} \times O(1) \, .
\end{equation}
The real part of 
$\PiT$ is $\sim \mD^2$, which is relevant for $(\omega-q) \sim
\mD^2/(\omega + q)$, that is, within $g^2 T$ of the light cone.
At timelike momentum but near 
the light cone, the leading order expansion for the real
and imaginary parts of $\PiT$, keeping the largest real and the largest
imaginary part, is
\begin{equation}
\label{eq:approx_PiT}
\PiT(\omega,q)\Big|_{\omega-q \ll \omega} \simeq
	\frac{\mD^2}{2} - i \frac{g^2 \nf}{12\pi}(\omega^2 - q^2)
	\left[1 - {\rm thermal\; part} \right]
\, .
\end{equation}
The thermal imaginary part is of the opposite sign as the vacuum
imaginary part, and is strictly smaller.
Its form is rather complicated even in the small $(\omega{-}q)$ limit.
Note that the imaginary part vanishes as the
momentum approaches the light cone, $q \rightarrow \omega$.  The
imaginary part vanishes in the standard HTL approximation to the
self-energy.  

The location of the mass shell, for $\omega \sim T \gg \mD$, is
\begin{equation}
\omega \simeq q + \frac{\mD^2}{4q} \, ,
\end{equation}
just as if the gauge boson had a mass squared of $\mD^2/2$.  The half
width of the mass shell is 
\begin{equation}
\Delta q_{1/2} \sim \frac{g^2 \nf \mD^2}{48 \pi \omega} \, ,
\end{equation}
which is very narrow, $O(g^4 \nf^2 T)$.  This gives a phase space
suppression to the resonance region.  The value of the matrix element
on resonance is
\begin{equation}
{\cal M}^2({\rm on \; shell}) \sim \frac{(\omega^2 - q^2)^2}
	{{\rm Im} \: \PiT^2} \sim \left( \frac{12 \pi}{g^2 \nf}
	\right)^2 \sim (g^2 \nf)^{-2} \, ,
\end{equation}
which is exactly enough to make up for the phase space suppression and
make the contribution to the scattering rate, from exchange of an
on-shell gauge boson, comparable to the integral over all exchanges of
off-shell gauge bosons.

An easier way of seeing this result is to consider the process
of on-shell gauge boson production, shown in Fig.~\ref{fig:2to1}.  This
process contains an explicit $g^2 \nf$, and it is also phase space
suppressed for kinematic reasons, by $O(\mD^2/T^2)$, which is $O(g^2
\nf)$.  Therefore it contributes $O(g^4 \nf^2)$ to \Eq{eq:Cmn}, which is
parametrically of the same order as the other scattering
processes, even when the coupling is taken small so the gauge boson
propagates close to the light cone.

We parenthetically remark that the ``formation timescale'' for the
collinear annihilation process just described is 
$\sim q/\mD^2 \sim 1/g^2 \nf T$.  This is shorter than the mean
time between scatterings for the fermions, $\sim 1 / g^2 T$, by a factor
of $1/\nf \ll 1$.  Therefore the annihilation process, although it
involves collinear physics, is not sensitive to scatterings of
the fermions at leading order; there is no Landau-Pomeranchuk-Migdal
suppression.  However, if we were speaking of QCD or QED {\em outside}
the large $\nf$ expansion, this would {\em not} be the case.  The
collinear pair annihilation to a gauge boson process would occur at a
rate of $O(1)$ importance to transport coefficients, but the rate itself
would get $O(1)$ corrections from scatterings on the incoming legs.
This has recently been discussed, in the context of photon emission from
the quark-gluon plasma, by Gelis et.~al.~\cite{Gelis}.

There are two strategies to compute the leading order in $g^2 \nf$
contribution of $s$ channel scattering to \Eq{eq:Cmn}.  One method is to
compute the $s$ channel scattering diagram, with no self-energy
insertion on the gauge boson propagator, and to add to it the rate of
the $f\bar f \leftrightarrow \gamma$ process just discussed, computed
using the corrected gauge boson dispersion relations and treating the
gauge boson as an external line.  This method means that we are allowing
gauge bosons to be external states in scattering processes, which means
that they are being included as kinetic degrees of freedom.  Such a
kinetic description of the fermions {\em and} gauge bosons is possible
at leading order in $g^2 \nf$ and should be distinguished from the
kinetic treatment of the fermions alone, which is accurate to $O(1/\nf)$
corrections as discussed above.

A kinetic treatment of both fermions and gauge bosons requires
considering the departure from equilibrium of the gauge bosons.  For
number diffusion this is trivial; the departure vanishes, as the gauge
bosons carry no fermionic number.  For shear viscosity it is not
trivial, though.  Therefore we turn to an
alternative, which gives the
same result.  It is to compute the rate 
of the $f\bar f \leftrightarrow f \bar f$ process
at finite $g^2 \nf$, using either the
full self-energy or the approximation of \Eq{eq:approx_PiT} above, 
and to extract
numerically the small $g^2 \nf$ limit.
The convergence to a small $g^2 \nf$ limit is very good, with
power suppressed corrections.

We emphasize, however, that it is {\em not} correct to compute the
leading order scattering contributions from $s$ channel exchange by
simply dropping the self-energy on the gauge boson line.  This misses
the contribution from resonant scattering through an on-shell gauge
boson.  It is also {\em not} correct to insert the HTL self-energy on
the gauge boson line.  The HTL approximation to the self-energies is
correct at leading order, not only at soft momentum $q,\omega \sim \mD$,
but at nearly lightlike momentum\footnote{We thank Tony Rebhan for
pointing this out to us.}
$\omega^2 - q^2 \sim \mD^2$, and this is the important regime.  However,
in the HTL approximation the self-energy has no imaginary part at
timelike momentum.
The presence of a small imaginary part is essential to
prevent a divergence in the scattering rate, so it is necessary here to
go beyond the HTL approximation and get the leading nonzero imaginary
part.

Numerically, the contribution of the on-resonance region is comparable
to the off-resonance $s$ channel region for the case of number
diffusion, and is a few times smaller for the case of shear viscosity.
The importance of {\em all} $s$ channel scattering processes, relative
to $t$ channel contributions, is numerically small for both transport
coefficients.  Throughout the range of coupling $g$ we have considered,
the $t$ channel scattering part of \Eq{eq:Cmn} dominates the $s$ channel
part by at least a factor of 3 for number diffusion and a factor of
20 for shear viscosity.  The domination grows larger for smaller
values of $g$.  The dominant role of $t$ channel scattering processes in
setting the shear viscosity is also found at leading log order
\cite{paper1}. 

\subsection{Leading order results}

\begin{figure}[p]
\vspace{-0.15in}
\centerline{\epsfxsize=6.2in\epsfbox{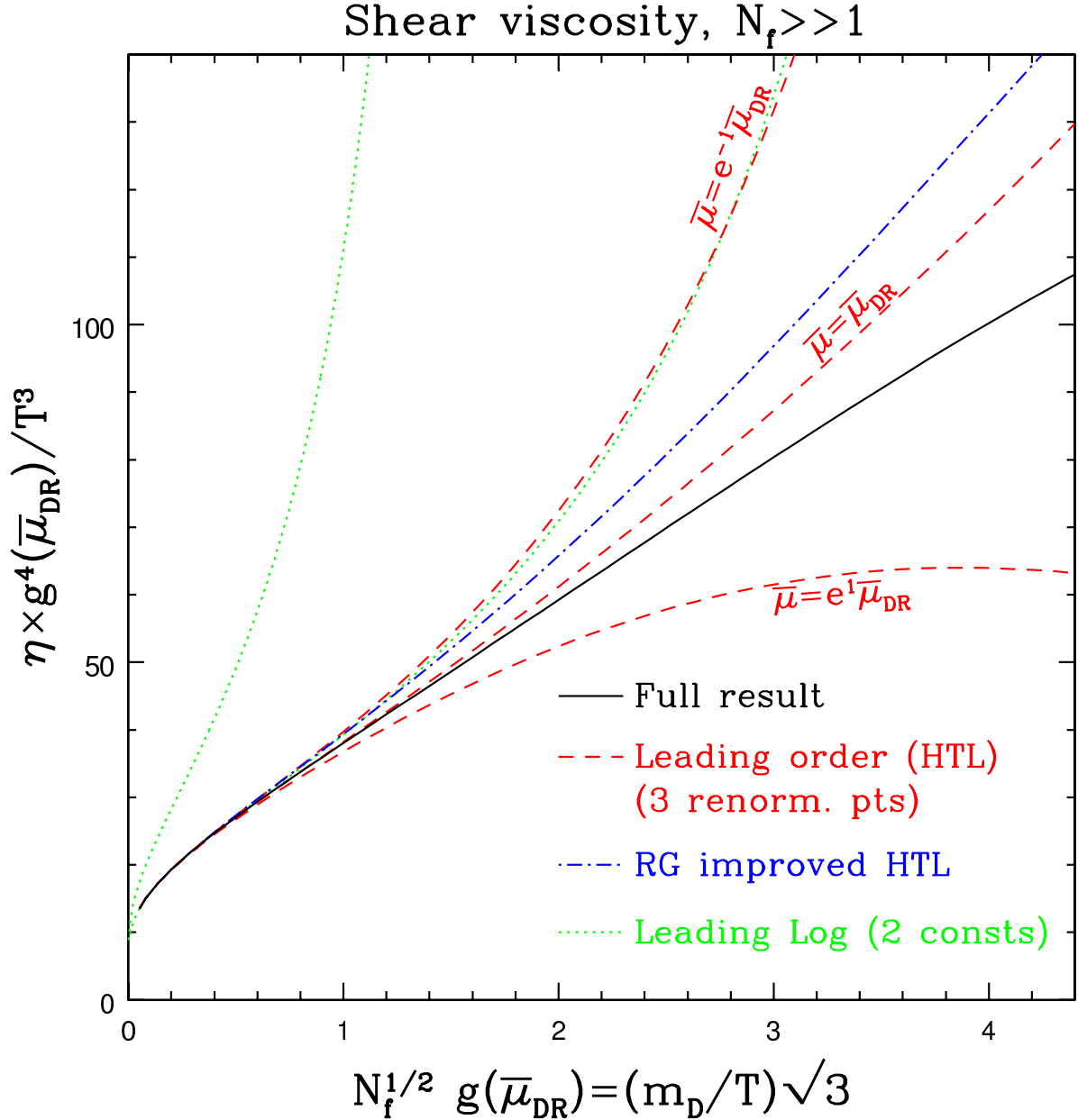}}
\vspace{0.15in}
\caption{\label{fig:compare1} Shear viscosity computed with various
approximations.  The leading order
calculation has an $O(g^2 \nf)$ ambiguity associated with the
renormalization point chosen for $g^2$; we show the result for three
choices of renormalization point, $\bar\mu=\bar\mu_{\rm DR}$, and $e$
and $1/e$ times this, and ``RG improved.''  
The leading log result is shown for two constants under the log.
The curve which grossly disagrees with all
others is using $O(1)=0$ in \Eq{eq:LL}, the other leading-log curve
picks the $O(1)$ constant to reproduce the leading order result at the
smallest value of $g \sqrt{\nf}$.}
\end{figure}

\begin{figure}[p]
\vspace{-0.15in}
\centerline{\epsfxsize=6.2in\epsfbox{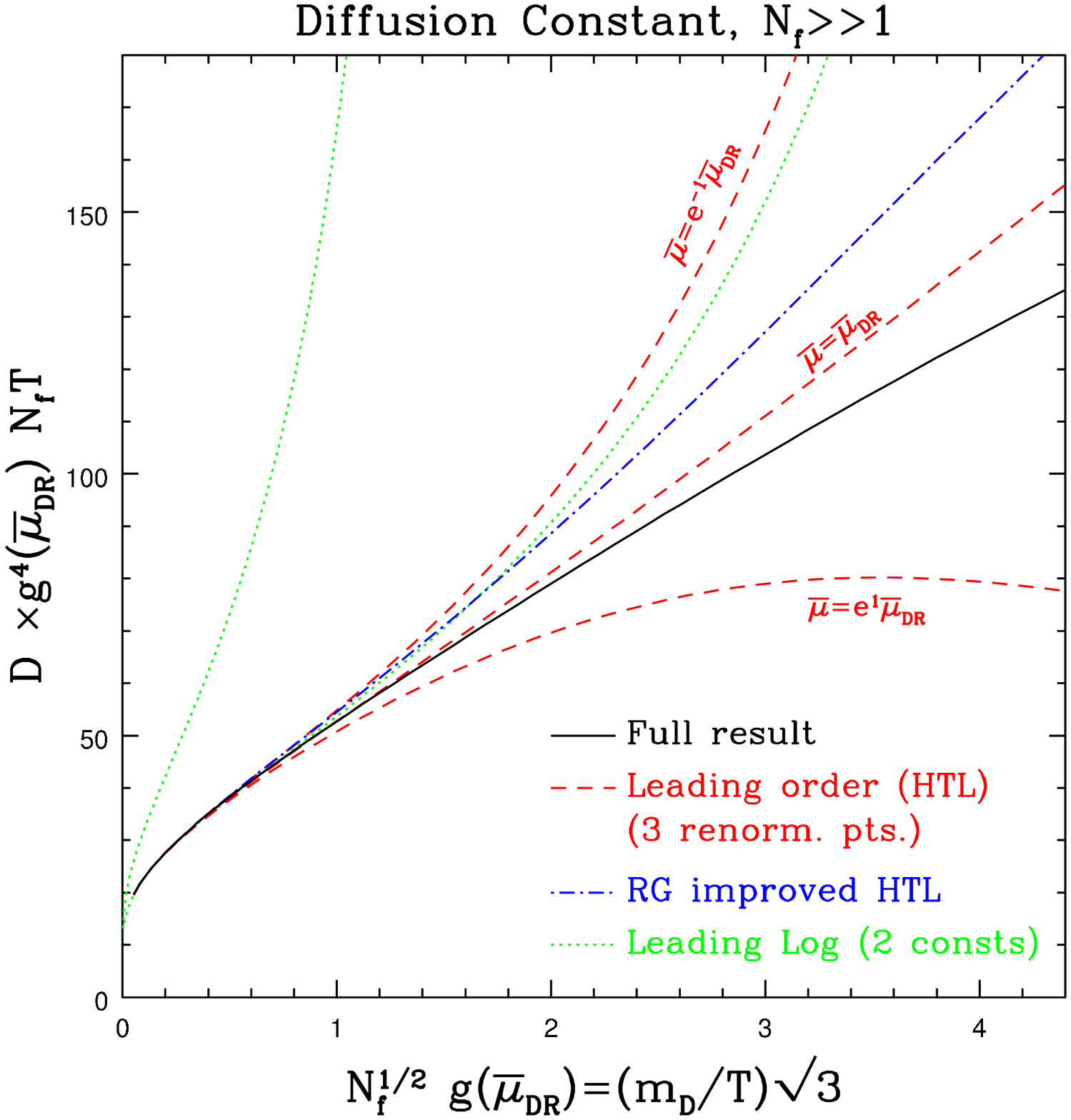}}
\vspace{0.15in}
\caption{\label{fig:compare2} Same as Fig.~\ref{fig:compare1}
but for flavor diffusion.}
\end{figure}

We compare the leading order calculation to the complete calculation in
Fig.~\ref{fig:compare1} and Fig.~\ref{fig:compare2}.  The errors due to
imprecision of numerical integrations are smaller than the line
thicknesses in the figures.  Note however that
there is a formally $O(g^2 \nf)$ ambiguity in the leading order
calculation.  The self-energy is taken to be only the HTL thermal one,
without the vacuum contribution.  Therefore, in the leading order
calculation, the value of $g^2$ does not vary with $q^\mu q_\mu = q^2 -
\omega^2$, but is fixed.  In the complete calculation the coupling is
the running coupling.  To compare the results of the leading order and
complete calculations we must decide at what renormalization point to
evaluate the running coupling constant, to insert as the fixed $g^2$ of
the leading order calculation.  This is a normal problem in gauge
theories; a calculation made to some order in a coupling constant
carries renormalization point dependence at one higher order in the
coupling.

There is a particularly natural choice for the renormalization point,
which is the one for which tree expressions give the correct infrared,
thermodynamic description of the gauge field physics.  This is the
``dimensional reduction'' value 
$\bar{\mu} = \pi e^{-\gamma_{\rm E}} T$ \cite{KLRS}.  We have
chosen this as a ``central'' value for the presentation of results, and
we illustrate the dependence on $\bar\mu$ by varying
renormalization point by a factor of $e$ on either side.

An alternative approach is to settle the renormalization point
dependence by including the real part of the 
(very simple) vacuum self-energy
contribution everywhere, in addition to the HTL approximation of the
thermal contribution.  This is equivalent to using
$g^2(\bar\mu = |\omega^2-q^2|)$ as the value of $g^2$ when the
exchange momentum is $(\omega,q)$--that is, allowing the gauge coupling
to run as the exchange momentum varies inside the integral in
\Eq{eq:Cmn}.  We therefore term this a ``renormalization group
improved'' leading order calculation.  It is not
correct beyond leading order in $g^2 \nf$, but it is free of
renormalization point ambiguities.  It is also shown in
Figs.~\ref{fig:compare1} and \ref{fig:compare2}.  The RG improved result
systematically lies above the $\bar\mu = \bar\mu_{\rm DR}$ value, and
turns out to be a worse approximation over the entire range considered.
Note the strikingly good performance of the $\bar\mu = \bar\mu_{\rm DR}$
renormalization point, leading-order result, especially at small
coupling.  We interpret this as the result of ``getting the
renormalization point right''; the leading thermal correction, left out
when we make the HTL approximation, cancels the vacuum running of the
coupling and effectively sets $g^2 \simeq g^2(\bar\mu_{\rm DR})$ in the soft
exchange momentum region.

The figures also show the result of the leading-log calculation.  There
is an even more severe ambiguity in the leading-log calculation than in
the leading order one; we do not know the $O(1)$ constant in
\Eq{eq:LL}.  We display two values.  The curve in each figure which
lies above all other curves is setting $O(1) = 0$ in \Eq{eq:LL}, while
the curve which tracks very closely the $\bar\mu = \bar\mu_{\rm DR}
e^{-1}$ leading order curve is the result if we choose the $O(1)$
constant to match the leading order calculation at the smallest value of
$\mD/T$ shown.  The ambiguity in the leading-log result is large even
where the coupling is weak, and the naive choice $O(1) = 0$ in
\Eq{eq:LL} gives horrible results.  However, what amounts to a ``next to
leading log'' calculation works surprisingly well, comparably to leading
order.

\section{Conclusions}
\label{sec:conclusion}

Computing transport coefficients, or any other quantity involving
infrared limits of correlation functions, is a difficult problem in the
context of relativistic gauge theories.  We have shown that, in a rather
special limit, this problem simplifies and the calculation, though still
difficult, is tractable.  This limit is the limit of a large number of
fermions, $\nf \gg 1$, expanding to leading nontrivial order in
$1/\nf$.  In this limit, nonabelian effects become unimportant.
Furthermore, the fermions enjoy nearly free dispersion relations, with
corrections (both dispersion and scattering corrections) suppressed by
$1/\nf$.  This allows a kinetic theory treatment of transport
coefficients, which we have presented.

This kinetic treatment is actually simpler than a kinetic treatment of
QED or QCD would be, if we intended to compute to leading order in $g$.
In particular, in large $\nf$ QED, there are no contributions at leading
order from Compton scattering and annihilation to gauge bosons, or from
interference diagrams or bremsstrahlung processes.  Nonabelian effects
are also irrelevant at leading order, so the QED and QCD calculations
differ only by group factors.  The structure of the
scattering integral is simple enough that it is possible to carry the
calculation out nonperturbatively in $g^2 \nf$, though this demands
some care in the treatment of the scattering matrix element.  In
particular, the hard thermal loop (HTL) approximation for the thermal
self-energy is inadequate, and a general expression for the thermal
self-energy is needed.

It is also possible to perform a leading order in $g^2 \nf$ 
calculation by making the
hard thermal loop (HTL) approximation.  This permits an interesting 
test of the quality of the HTL approximation.  Because the physics of
this model is missing some complications present in QED and QCD (such as
bremsstrahlung), we expect the HTL approximation to work, if anything,
better in this model than in realistic QED or QCD. In fact, corrections
to the HTL leading order calculation which we made here are
parametrically $O(g^2 \nf)$, while in full QED or QCD we believe that
gauge boson emission processes and nonabelian effects (QCD only) will
introduce $O(g)$ corrections.

Our results for the comparison between the complete solution and the HTL
approximation are presented in Fig.~\ref{fig:compare1} and
Fig.~\ref{fig:compare2}, which constitute the main results of this paper.
They show that, for $g^2\nf \leq 1$, the leading order treatment is
accurate to a few percent.  
This is good news for QED, where the coupling is indeed
small.  For instance, between the electron and muon mass scales, the QED
coupling is $g \sqrt{\nf} \simeq .30$.  At this coupling the difference
between the HTL and full treatments is less than $0.3\%$.  Even at this
coupling, however, a calculation to leading order in the {\em logarithm}
of the coupling is of very little value.

On the other hand, as the coupling grows larger, the leading order
treatment becomes much less 
reliable and more renormalization point dependent.  In the theory we
have looked at, the size of the renormalization point dependence of the
leading order calculation and the size of its error (its difference from
the exact result) are comparable.
However, the ``RG improved'' calculation shows that the error is
probably not due to a poor choice of renormalization point, but is a
failure of the HTL approximation to correctly represent the thermal part
of the self-energy.  This is mixed news for QCD.  

We can estimate how accurate a leading order calculation
of transport coefficients in QCD would be by seeing how accurate the
large $\nf$ results are at a value of the Debye mass $\mD/T$
equal to the relevant QCD value.  First consider
QCD for $T$ at the electroweak scale, $T \sim 100$GeV.  Taking $\alphas
= 0.11$, $\nf=6$ and using \Eq{eq:mDQCD}, we find this corresponds to $g
\sqrt{\nf} \simeq 2.9$.  For this value, the renormalization point
uncertainty in the leading order value
is about $30\%$.  This is not so bad, though in full QCD the
actual error will likely be larger than this.

The correction is much more severe at $T \sim $1GeV, where the coupling
is much larger.  This value of $T$ gives a $g$ roughly equal to
the right hand edge of the figures, where the renormalization point
sensitivity of the leading order calculation is at least
a factor of 2.  Therefore, at
temperatures which may realistically be obtained in heavy ion collisions
in the foreseeable future, a full leading order calculation is at best a
factor of 2 estimate.  

\section*{Acknowledgements}

I thank Dietrich B\"{o}deker, Francois Gelis, Emil Mottola, 
Tony Rebhan, and Larry
Yaffe for helpful conversations.  I particularly thank Tony Rebhan, for
pointing out an error (see \Eq{eq:53}) more than a year after the original
publication; correcting it reduced the difference between leading-order
and exact results.  This work 
was partially supported by the DOE under contract DE-FGO3-96-ER40956.

\appendix

\section{Integration variables}
\label{app:phasespace}

Both $s$ and $t$ channel scattering processes must be computed in the
main text.  Different simplifications of the phase space integration
in \Eq{eq:Cmn} prove most convenient for $s$ and $t$ channel
scattering processes.  It is convenient
to have, as integration variables, the magnitudes of the spatial and
temporal components of the momentum carried by the gauge boson, since it
is by far easiest to express $|{\cal M}|^2$ in terms of these.
Here we present both choices for phase space
parameterization we have found useful.  Note that we always take the
external momenta to be strictly lightlike (massless dispersion
relations), meaning $p_0 = |\p| \equiv p$, which is justified at leading
order in $1/\nf$.

\subsection{$t$ channel exchange}
\label{t_phase}

First consider the case were $\p'-\p$ is the exchange momentum.
Then in the collision integral , \Eq{eq:Cmn}, it is 
convenient to use the spatial $\delta$ function to perform the
$\k'$ integration, and to shift the $\p'$ integration into an
integration over $\p'{-}\p \equiv \q$.
We may write the angular integrals
in spherical coordinates with $\q$ as the $z$ axis and choose the $x$
axis so $\p$ lies in the $x$-$z$ plane.
This yields
\begin{eqnarray}
&& \beta^3 \int \frac{d^3 \p d^3 \k d^3 \p' d^3 \k'}{(2\pi)^{12} 16 pkp'k'}
	(2\pi)^4 \delta^3(\p+\k-\p'-\k') \delta(p+k-p'-k')
\nonumber \\ & = & 
    \frac{\beta^3}{2^9 \pi^6}
    \int_{0}^{\infty} \! \! \! q^2 dq \> p^2 dp \> k^2 dk 
	\int_{-1}^{1} \! \! d \cos\theta_{pq} \> 
	d\cos\theta_{kq}
	\int_0^{2\pi} \! \! d\phi
	\frac{1}{p\,k\,p'\,k'}
	\delta(p{+}k{-}p' {-} k') \, ,
\end{eqnarray}
where $q=|\q|$, and $p'$ and $k'$ become dependent variables,
$p' \equiv |\q+\p|$ and $k' \equiv |\k-\q|$.
$\phi$ is the azimuthal angle of $\k$ (and $\k'$)
[{\em i.e.}, the angle between the $\p$-$\q$ plane and the $\k$-$\q$ plane],
and $\theta_{pq}$ is the plasma frame angle between $\p$ and $\q$,
$\cos\theta_{pq} \equiv \hat\p \cdot \hat\q$, {\em etc}.

Following Baym {\it et~al.}~\cite{BMPRa}, it is convenient to
introduce a dummy
integration variable $\omega$, defined by a $\delta$ function to equal
the energy transfer $p' - p$,
so that
\begin{equation}
    \delta(p+k-p' - k')
    =
    \int_{-\infty}^\infty d\omega \>
    \delta(\omega + p - p') \, \delta(\omega-k+k') \, .
\end{equation}
Evaluating $p'=|\p+\q|$
in terms of $p$, $q$, and $\cos \theta_{pq}$, and 
defining $t = \omega^2 - q^2$ (which is the usual Mandelstam variable),
one finds
\begin{eqnarray}
    \delta(\omega+p-p')
    &=&
    \frac{p'}{pq} \>
    \delta\biggl( \cos \theta_{pq} - \frac{\omega}{q} 
	- \frac{t}{2pq} \biggr) \Theta(\omega + p )\,,
\\
    \delta(\omega-k+k')
    &=&
    \frac{k'}{kq} \>
    \delta\biggl( \cos \theta_{kq} - \frac{\omega}{q} 
	+ \frac{t}{2kq} \biggr) \Theta ( k - \omega ) \,.
\end{eqnarray}
Here $\Theta$ is the step function.
The $\cos\theta$ integrals may now be trivially performed and yield 1
provided $p > \half(q-\omega)$, $k>\half(q+\omega)$, and
$|\omega|<q$; otherwise the argument of a $\delta$ function has no
zero for any $|\cos \theta| \leq 1$.
The integration range becomes
\begin{equation}
    \frac{\beta^3}{2^{9} \pi^6}
    \int_0^{\infty} \! \! dq
    \int_{-q}^q \! d\omega 
    \int_{\frac{q-\omega}{2}}^\infty dp 
    \int_{\frac{q+\omega}{2}}^\infty dk
    \int_0^{2\pi} \! \! d\phi \, ,
\label{eq:int_t}
\end{equation}
with $p'=p{+}\omega$, $k'=k{-}\omega$.
For evaluating the final factor of \Eq{eq:Cmn},
note that
\begin{equation}
    I_\ij(\hat\p) \, I_\ij(\hat\k)
    =
	P_\ell(\cos\theta_{pk}) \,,
\la{eq:contract_I}
\end{equation}
where $P_\ell$ is the $\ell$'th Legendre polynomial, and $\ell=1$ if
there is one index (diffusion) and $\ell = 2$ if there are two indices
(shear viscosity).  
We will therefore need expressions for the angles between all species,
as well as the remaining Mandelstam variables $s$ and $u$, which may 
appear in ${\cal M}^2$.
They are
\begin {eqnarray}
\la{eq:s_equals}
    s &=& \frac{-t}{2q^2}
	\left\{
	\left[(p+p')(k+k')+q^2 \right]
	    - \cos\phi \>
		\sqrt{\textstyle
		\left(4pp'+t \right)\left(4kk'+t \right) } \,
	\right\}
	\, , \\
u & = & -t -s \, ,
\label{eq:u_equals}
\end{eqnarray}
and
\begin{eqnarray}
\cos \theta_{pq}\: & = & \frac{\omega}{q} + \frac{t}{2pq} 
	\, , \hspace{1.07in} \;
\cos \theta_{p'q} \: = \frac{\omega}{q} - \frac{t}{2p'q} \, , \nonumber \\
\cos \theta_{kq} \: & = & \frac{\omega}{q} - \frac{t}{2kq} 
	\, , \hspace{1.1in}
\cos \theta_{k'q}\: = \frac{\omega}{q} + \frac{t}{2k'q} \, , \nonumber \\
\cos \theta_{pp'} & = & 1 + \frac{t}{2pp'} \, ,\hspace{1.12in}
\cos \theta_{kk'} \: = 1 + \frac{t}{2kk'} \, , \nonumber \\
\cos \theta_{pk'} & = & 1 + \frac{u}{2pk'}
	\, , \hspace{1.115in}
\cos \theta_{p'k} \: = 1 + \frac{u}{2p'k}
	\, , \nonumber \\
\cos \theta_{pk} \: & = & 1 - \frac{s}{2pk}
	\, , \hspace{1.15in}
\cos \theta_{p'k'} = 1 - \frac{s}{2p'k'}
	\, .
\label{eq:angles}
\end {eqnarray}
The $\phi$ integration can be done analytically in all cases
in this paper, but the other integrals cannot.

\subsection{$s$ channel exchange}
\label{s_phase}

When the exchange gauge boson carries momentum $\p+\k$, it is
convenient to use a different parameterization of phase space.  Again the 
spatial $\delta$ function in Eq.~(\ref{eq:Cmn}) is used to perform the
$\k'$ integration, but now the $k$ integration is shifted to an integral 
over $\q \equiv \p + \k$, the total incoming spatial momentum.
Again we use spherical coordinates, with $\q$ on the $z$ axis and $\p$
in the $x$-$z$ plane.  
In these variables, the integration measure becomes
\begin{eqnarray}
&& \beta^3 \int \frac{d^3 \p d^3 \k d^3 \p' d^3 \k'}{(2\pi)^{12} 16 pkp'k'}
	(2\pi)^4 \delta^3(\p+\k-\p'-\k') \delta(p+k-p'-k')
\nonumber \\ & = & 
    \frac{\beta^3}{2^9 \pi^6}
    \int_{0}^{\infty} \! \! \! q^2 dq \> p^2 dp \> {p'}^2 dp' 
	\int_{-1}^{1} \! \! d \cos\theta_{pq} \> 
	d\cos\theta_{p'q}
	\int_0^{2\pi} \! \! d\phi
	\frac{1}{p\,k\,p'\,k'} \>
	\delta(p{+}k{-}p' {-} k') \, ,
\label{int_s}
\end{eqnarray}
where $k=|\q{-}\p|$ and $k'=|\q{-}\p'|$, and $\phi$ is the azimuthal angle
of $\p'$, meaning the angle between the $\q,\p$ plane and the $\q,\p'$
plane. 

Next, we re-write the energy delta function in terms of $\omega$, the 
total energy;
\begin{equation}
\delta(p+k-p'-k') = \int_{0}^{\infty} d\omega \> \delta(\omega - p - k)
	\, \delta ( \omega - p' - k' ) \, .
\end{equation}
In terms of the remaining integration variables, and defining $s =
\omega^2 - q^2$ (which again is the Mandelstam variable), one finds
\begin{eqnarray}
\delta(\omega - p - k) & = & \frac{k}{pq} \: \, \delta \left( 
	\cos \theta_{pq} \: - \frac{\omega}{q} 
	+ \frac{s}{2pq} \: \right) \Theta(\omega - p \:) \, , \\
\delta(\omega - p' - k') & = & \frac{k'}{p'q} \, \delta \left( 
	\cos \theta_{p'q} - \frac{\omega}{q} 
	+ \frac{s}{2p'q} \right) \Theta(\omega - p') \, ,
\end{eqnarray}
which performs the integrals over $\cos \theta_{pq}$ and $\cos
\theta_{p'q}$.  There is only a solution provided $q < \omega$, $|2p -
\omega| < q$, and $|2 p' - \omega| < q$.
The integration measure becomes
\begin{equation}
	\frac{\beta^3}{2^9 \pi^6}
	\int_0^\infty \! \! d\omega \int_0^\omega \! dq 
	\int_{\omega-q \over 2}^{\omega+q \over 2} \! \! dp
	\int_{\omega-q \over 2}^{\omega+q \over 2} \! \! dp'
	\int_0^{2\pi} \! \! d\phi	 \, ,
\label{eq:int_s}
\end{equation}
with $k=\omega{-}p$, $k'=\omega{-}p'$.  The Mandelstam variables are
\begin{eqnarray}
t & = & \frac{s}{2q^2} \left\{ \left[ (p-k)(p'-k') - q^2 \right] 
	+ \cos \phi \sqrt{(4pk-s)(4p'k'-s)} 
	\right\} \, , \\
u & = & -s -t \, .
\end{eqnarray}
The angles with respect to $\q$ are
\begin{eqnarray}
\cos \theta_{pq} & = & \frac{\omega}{q} - \frac{s}{2pq}
	\, , \hspace{1.08in}
\cos \theta_{p'q} = \frac{\omega}{q} - \frac{s}{2p'q}
	\, , \nonumber \\
\cos \theta_{kq} & = & \frac{\omega}{q} - \frac{s}{2kq}
	\, , \hspace{1.08in}
\cos \theta_{k'q} = \frac{\omega}{q} - \frac{s}{2k'q}
	\, , 
\end{eqnarray}
and the remaining angles are as in \Eq{eq:angles}.

Again the $\phi$ integral can be performed analytically, but
the other integrals cannot.

\section{Matrix elements}
\label{app:Msq}

In this appendix we present the matrix elements for $s$ and $t$ channel
gauge boson exchange.  These must be calculated {\em without} making any
expansion in $q \ll p$ or in small self-energy; for our purpose the
self-energy must be taken to be an $O(1)$ correction and all momenta
must be considered to be the same order.

The quantity ``[$t^2 + u^2$$/s^2$]'', introduced as a shorthand in
\Eq{eq:schann}, technically means
\begin{equation}
\frac{1}{8} D^{\rm ret}_{\mu \alpha} D^{\rm adv}_{\nu \beta} 
	{\rm Tr} ( \nott{p} \gamma^\mu 
	\nott{k} \gamma^\nu ) 
	{\rm Tr} ( \nott{p}' \gamma^\alpha 
	\nott{k}' \gamma^\beta ) \, .
\label{eq:meaning1}
\end{equation}
Following the previous appendix we denote the plasma frame frequency and
momentum carried by the gauge boson propagator as $\omega$ and $q$.
The retarded gauge boson propagator $D^{\rm ret}_{\mu \nu}$ is 
most conveniently expressed in
Coulomb gauge--the result for \Eq{eq:meaning1} being gauge invariant:
\begin{eqnarray}
D^{\rm ret}_{00}(\omega,\q) & = & \frac{1}{q^2 
	- \PiL^{\rm ret}(\omega,q)} \, , \\
D^{\rm ret}_{ij}(\omega,\q) & = & \frac{\delta_{ij} - \hat\q_i \hat\q_j}
	{q^2 - \omega^2 + \PiT^{\rm ret}(\omega,q)} \, , \\
D^{\rm ret}_{0i}(\omega,\q) & = & D_{i0}(\omega,\q) \: = \: 0 \, . \\
\end{eqnarray}
The sign of $D^{\rm ret}_{ij}$ is opposite the most common convention
because we use a $({-}{+}{+}{+})$ metric; but we have chosen the sign of
$\PiT$ to correspond to common usage.
The advanced propagator is given by complex conjugating these
expressions.  In what follows we drop the [ret] superscript from the
self-energies.  

\pagebreak

It will be convenient for our purposes to define nonstandard
normalization self-energies
\begin{equation}
\tilde \PiL \equiv - \PiL \, , \qquad
\tilde \PiT \equiv \frac{q^2}{q^2 - \omega^2} \PiT \, .
\end{equation}
In terms of these, evaluating \Eq{eq:meaning1} gives
\begin{eqnarray}
2 \ttuuss & = &
	\frac{1}{|q^2 + \tilde \PiL|^2} (4pk{-}s)(4p'k'{-}s) \nonumber \\ &&
	+\frac{2}{\left| (q^2 + \tilde \PiL^*)(q^2 + \tilde \PiT) \right|}
	 (p{-}k)(p'{-}k') \sqrt{(4pk{-}s)(4p'k'{-}s)} 
	\, \cos \phi \nonumber \\ &&
	+ \frac{1}{|q^2 + \tilde \PiT|^2} \left(
	(4pk{-}s)(4p'k'{-}s)\, \cos^2 \phi + q^2(2\omega^2 - 4pk -4p'k') 
	\right) \, .
\label{eq:s_M}
\end{eqnarray}
The analogous result for $t$ channel exchange is
\begin{eqnarray}
2 \ssuutt & = &
	\frac{1}{|q^2 + \tilde \PiL|^2} (4pp'{+}t)(4kk'{+}t) \nonumber \\ &&
	-\frac{2}{\left| (q^2 + \tilde \PiL^*)(q^2 + \tilde \PiT) \right|}
	 (p{+}p')(k{+}k') \sqrt{(4pp'{+}t)(4kk'{+}t)} 
	\, \cos \phi \nonumber \\ &&
	+ \frac{1}{|q^2 + \tilde \PiT|^2} \left(
	(4pp'{+}t)(4kk'{+}t)\, \cos^2 \phi + q^2(2\omega^2 + 4pp' +4kk') 
	\right) \, .
\label{eq:t_M}
\end{eqnarray}
The variables in these expressions, such as $\phi$, $s$, $t$, $p$, etc
are defined in Appendix \ref{app:phasespace}.  In particular the
definition of $\phi$ in \Eq{eq:s_M} is the angle between the $\q,\p$
plane and the $\q,\p'$ plane, whereas in \Eq{eq:t_M} it is the angle
between the $\q,\p$ and $\q,\k$ planes.

Complete expressions for $\PiL$ and $\PiT$ were obtained
by Weldon \cite{Weldon1}, and the usual expressions for the hard thermal
loop limit of the self-energies were extracted as a particular limit.
The vacuum parts are simple%
\footnote
    {%
    The earier (and published!) version is missing the $5/3$ term here,
    an error which arose from an error in Weldon's paper \cite{Weldon1} (unless
    his dimensionally regularized renormalization point $\sigma$ 
    in his Eq.~(A2) is meant to be $\bar\mu \, e^{5/6}$).
    I thank Tony Rebhan for pointing out this error, which is rather
    significant at the largest values of $g^2 \nf$ considered here.%
    }:
\begin{equation}
\label{eq:53}
\tilde \Pi_{\rm T,\; vac} = \tilde \Pi_{\rm L,\; vac}
	= \frac{g^2 \nf}{12\pi^2} q^2 \left[ -\log\left(
	\frac{|\omega^2 -q^2|}{\bar\mu^2} \right) + \frac{5}{3}
	+ i \pi \Theta(\omega^2-q^2) \right] \, .
\end{equation}
The imaginary part exists for timelike momenta and represents the vacuum
rate of decay into a fermion pair.  The full self-energy is the sum of
a vacuum and a thermal part.  The thermal parts of the self-energies are
\begin{equation}
\tilde \Pi_{\rm L,\; th} = \frac{g^2 \nf T^2}{3} H(\omega,q) \, , \qquad
\tilde \Pi_{\rm T,\; th} = \frac{g^2 \nf T^2}{3} 
	\left( - \frac{1}{2} H(\omega,q) 
	+ \frac{q^2}{2(q^2-\omega^2)} G(\omega,q) \right) \, ,
\end{equation}
where $H(\omega,q)$ and $G(\omega,q)$ are,\footnote
	{
	Our notation does not quite agree with \pcite{Weldon1}; we 
	separate out the Debye mass squared $g^2 \nf T^2 / 3$ from
	the definitions of $G$ and $H$, which are then pure numbers
	and have particularly simple small $\omega,q$ limits.
	} 
using the shorthand 
$2 \omega_+ \equiv \omega{+}q$ and 
$2 \omega_- \equiv \omega{-}q$,
\begin{eqnarray}
\Re \: G(\omega,q) & = & \frac{3}{\pi^2 T^2} \int_0^\infty \! \! \! 
	dk \: f_0(k) \left( 4k + \frac{q^2{-}\omega^2}{2q}
	\ln \left[ \frac{ (k{+}\omega_-)(k{-}\omega_+)}
	{(k{+}\omega_+)(k{-}\omega_-)} \right] \right) \, , \nonumber \\
\Re \: H(\omega,q) & = & \frac{3}{\pi^2 T^2} \int_0^\infty \! \! \! 
	dk \: f_0(k) \Bigg( 2k - \frac{4k^2{+}\omega^2{-}q^2}{4q}
	\ln \left[ \frac{ (k{+}\omega_-)(k{-}\omega_+)}
	{(k{+}\omega_+)(k{-}\omega_-)} \right] \nonumber \\ && \qquad \qquad
	- \frac{2k\omega}{q} \ln \frac{\omega_+}{\omega_-}
	- \frac{k\omega}{q} \ln \left[ \frac{(k{+}\omega_-)(k{-}\omega_-)}
	{(k{+}\omega_+)(k{-}\omega_+)} \right] \Bigg) \, , \nonumber \\
\Im \: G(\omega,q) & = & \frac{3}{\pi T^2} \int_0^\infty \! \! \! 
	dk \: f_0(k) \frac{q^2{-}\omega^2}{2q}
	\Big[ \Theta(k{+}\omega_+)\Theta(-k{-}\omega_-)	
	- \Theta(-k{+}\omega_+)\Theta(k{-}\omega_-) \Big] \, , \nonumber \\
\Im \: H(\omega,q) & = & \frac{3}{\pi T^2} \int_0^\infty \! \! \! 
	dk \: f_0(k) \Bigg[ - \frac{(2k{+}\omega)^2-q^2}{4q}
	\Theta(k{+}\omega_+)\Theta(-k{-}\omega_-) 
	\nonumber \\ && \hspace{1.36in}
	+ \frac{(2k{-}\omega)^2 -q^2}{4q} \
	\Theta(k{-}\omega_-)\Theta(-k{+}\omega_+)
	\nonumber \\ && \hspace{1.36in}
	+ \frac{2k\omega}{q} \Theta(\omega_+) \Theta(-\omega_-) \Bigg]
	\, .
\end{eqnarray}
The hard thermal loop limit means taking $\omega \ll T$, $q \ll T$, and
extracting the nonvanishing part.  The values of these integrals, in
this limit, depend only on $\omega/q$ and are
\begin{eqnarray}
H_{\rm HTL}(\omega/q) & = & 1 - \frac{\omega}{2q} 
	\ln \frac{\omega_+}{\omega_-} + i \frac{\pi \omega}{2q}
	\Theta(\omega_+)\Theta(-\omega_-) \, , \nonumber \\
G_{\rm HTL}(\omega/q) & = & 1 \, .
\label{eq:HTL}
\end{eqnarray}
However at general
$\omega/T$, $q/T$, only a few of the integrals can be performed
analytically and some have to be done numerically.  However, all
required integrals can be done quickly and to high precision by
numerical quadratures integration, and their evaluation does not limit
either the accuracy or speed of any subsequent integrations.

Note that, for timelike momenta, the thermal contributions to the 
imaginary parts of the self-energies are of opposite sign as the vacuum
parts, and represent Pauli blocking of pair production.  The thermal
contribution to the imaginary part at spacelike momenta, absent in the
vacuum theory, represents Landau damping.  In the HTL limit only the
imaginary part arising from $\Theta(\omega_+) \Theta(-\omega_-)$
contributes.  All real parts are even in $\omega$ and all imaginary
parts are odd in $\omega$.


\begin {references}

\bibitem{Keldysh}
L.~V.~Keldysh,
Zh.\ Eksp.\ Teor.\ Fiz.\ {\bf 47}, 1515 (1964).

\bibitem {HTL}
E.~Braaten and R.~D.~Pisarski,
Nucl.\ Phys.\  {\bf B337} (1990) 569;
J. Frenkel and J. Taylor, Nucl. Phys. {\bf B334}, 199 (1990);
J. Taylor and S. Wong, Nucl. Phys. {\bf B346}, 115 (1990). 

\bibitem{damprate}
E.~Braaten and R.~D.~Pisarski,
Phys.\ Rev.\ D {\bf 42}, 2156 (1990);
E.~Braaten and R.~D.~Pisarski,
Phys.\ Rev.\ D {\bf 46}, 1829 (1992);
T.~S.~Biro and M.~H.~Thoma,
Phys.\ Rev.\ D {\bf 54}, 3465 (1996)
[hep-ph/9603339];

\bibitem{energyloss}
E.~Braaten and M.~H.~Thoma,
Phys.\ Rev.\ D {\bf 44}, 1298 (1991).

\bibitem{bodeker}
D.~Bodeker,
Phys.\ Lett.\ B {\bf 426}, 351 (1998)
[hep-ph/9801430];
Nucl.\ Phys.\ B {\bf 566}, 402 (2000)
[hep-ph/9903478];
Nucl.\ Phys.\ B {\bf 559}, 502 (1999)
[hep-ph/9905239].

\bibitem {Jeon}
S.~Jeon,
Phys.\ Rev.\  {\bf D52}, 3591 (1995)
[hep-ph/9409250].

\bibitem {HosoyaKajantie}
A.~Hosoya and K.~Kajantie,
Nucl.\ Phys.\  {\bf B250}, 666 (1985).

\bibitem{Hosoya_and_co}
A.~Hosoya, M.~Sakagami and M.~Takao,
Annals Phys.\  {\bf 154}, 229 (1984).

\bibitem {relax1}
S.~Chakrabarty,
Pramana {\bf 25}, 673 (1985).

\bibitem {relax2}
W.~Czy\.{z} and W.~Florkowski,
Acta Phys.\ Polon.\  {\bf B17}, 819 (1986).

\bibitem {relax3}
D.~W.~von Oertzen,
Phys.\ Lett.\  {\bf B280}, 103 (1992).

\bibitem {relax4}
M.~H.~Thoma,
Phys.\ Lett.\  {\bf B269}, 144 (1991).

\bibitem {BMPRa}
G.~Baym, H.~Monien, C.~J.~Pethick and D.~G.~Ravenhall,
Phys.\ Rev.\ Lett.\  {\bf 64}, 1867 (1990);
Nucl.\ Phys.\  {\bf A525}, 415C (1991).
%
%

\bibitem {Heiselberg}
H.~Heiselberg,
Phys.\ Rev.\  {\bf D49}, 4739 (1994)
[hep-ph/9401309].

\bibitem {Heiselberg_diff}
H.~Heiselberg,
Phys.\ Rev.\ Lett.\  {\bf 72}, 3013 (1994)
[hep-ph/9401317].

\bibitem {BaymHeiselberg}
G.~Baym and H.~Heiselberg,
Phys.\ Rev.\  {\bf D56}, 5254 (1997)
[astro-ph/9704214].

\bibitem {JPT1}
M.~Joyce, T.~Prokopec and N.~Turok,
Phys.\ Rev.\  {\bf D53}, 2930 (1996)
[hep-ph/9410281].

\bibitem {MooreProkopec}
G.~D.~Moore and T.~Prokopec,
Phys.\ Rev.\  {\bf D52}, 7182 (1995)
[hep-ph/9506475].

\bibitem {JPT2}
M.~Joyce, T.~Prokopec and N.~Turok,
Phys.\ Rev.\  {\bf D53}, 2958 (1996)
[hep-ph/9410282].

\bibitem{paper1}
P.~Arnold, G.~D.~Moore and L.~G.~Yaffe,
JHEP{\bf 0011}, 001 (2000)
[hep-ph/0010177].

\bibitem{BraatenNieto}
E.~Braaten and A.~Nieto,
Phys.\ Rev.\ D {\bf 53}, 3421 (1996)
[hep-ph/9510408].

\bibitem{KadanoffBaym} L.~P.~Kadanoff and G. Baym, 
``Quantum Statistical Mechanics,''
Benjamin, New York (1962).

\bibitem{CalzettaHu}
E.~Calzetta and B.~L.~Hu,
Phys.\ Rev.\  {\bf D37}, 2878 (1988);
E.~A.~Calzetta, B.~L.~Hu and S.~A.~Ramsey,
Phys.\ Rev.\  {\bf D61}, 125013 (2000)
[hep-ph/9910334].

\bibitem{deGroot}
S.~R.~De Groot, W.~A.~Van Leeuwen and C.~G.~Van Weert,
``Relativistic Kinetic Theory. Principles And Applications,''
{\it  Amsterdam, Netherlands: North-holland ( 1980) 417p}.

\bibitem{KLRS}
K.~Kajantie, M.~Laine, K.~Rummukainen and M.~Shaposhnikov,
Nucl.\ Phys.\ B {\bf 458}, 90 (1996)
[hep-ph/9508379].
 
\bibitem{Gelis}
P.~Aurenche, F.~Gelis, R.~Kobes and H.~Zaraket,
Phys.\ Rev.\ D {\bf 60}, 076002 (1999)
[hep-ph/9903307];
P.~Aurenche, F.~Gelis and H.~Zaraket,
Phys.\ Rev.\ D {\bf 61}, 116001 (2000)
[hep-ph/9911367];
P.~Aurenche, F.~Gelis and H.~Zaraket,
Phys.\ Rev.\ D {\bf 62}, 096012 (2000)
[hep-ph/0003326].

\bibitem{Weldon1}
H.~A.~Weldon,
Phys.\ Rev.\  {\bf D26}, 1394 (1982).

\end {references}
\end{document}